\documentclass[aps,prd,
eqsecnum,showpacs,nofootinbib]{revtex4}

\usepackage{amsmath,amsfonts,amssymb,color,graphicx,latexsym,theorem,mathrsfs
}

\newcommand{\ma}[1]{\mbox{$\mathcal{#1}$}}
\newcommand{\mas}[1]{\mbox{$\mathscr{#1}$}}
\newcommand{\D}{{\rm d}}
\newcommand{\we}{\wedge}

\newcommand{\I}{{\rm i }}

\begin{document}

\title{
On the Bogomol'nyi bound in Einstein-Maxwell-dilaton gravity
}

\author{Masato Nozawa}
\email{nozawam_at_post.kek.jp}


\address{
Theory Center, KEK, Tsukuba 305-0801, Japan
}

\date{\today}

\begin{abstract} 
It has been shown that the 4-dimensional Einstein-Maxwell-dilaton
theory allows a Bogomol'nyi-type inequality for an arbitrary  
dilaton coupling constant $\alpha $, and that 
the bound is saturated if and only if the (asymptotically flat) spacetime admits   
a nontrivial spinor satisfying the gravitino and the dilatino Killing spinor
equations. The present paper revisits this issue and argues that the
dilatino equation fails to ensure the dilaton field equation unless the solution is purely electric/magnetic, or 
the dilaton coupling constant is given by $\alpha=0, \sqrt 3$, corresponding
to the Brans-Dicke-Maxwell theory and the 
Kaluza-Klein reduction of 5-dimensional vacuum gravity, respectively.  
A systematic classification of the supersymmetric solutions reveals that
the solution can be rotating if and only if the solution is dyonic or 
the coupling constant is given by $\alpha=0, \sqrt 3$.
This implies that the theory with $\alpha \ne 0, \sqrt 3$
cannot be embedded into supergravity except for the static truncation. 
Physical properties of supersymmetric solutions are explored
from various points of view.  
\end{abstract}


\maketitle


\section{Introduction}

Effective gravitational theories obtained via the 
Kaluza-Klein paradigm have attracted much attention and have continued
to give a lot of physical insights into unified theories.  
In the low energy limit of string theory, one recovers  
Einstein's gravity with a dilatonic scalar field arising from
dimensional reduction. A dilaton field   
naturally couples to several gauge fields with various ranks, and    
its coupling constant depends on the underlying
theory and the dimension of an internal space. 
A variety of physical phenomena may be influenced by a dilaton field. 
An illuminating example is the asymptotically flat, static and
spherically symmetric black hole solutions to the
Einstein-Maxwell-dilaton 
system~\cite{Gibbons:1982ih,GM,Garfinkle:1990qj}. 
They exhibit novel aspects compared to the Reissner-Nordstr\"om
solution in the Einstein-Maxwell theory:  
the inner ``horizon'' of a black hole is a spacelike singularity 
and the Hawking temperature in the ``extreme'' case can be
non-vanishing. These properties alter significantly 
the spacetime structure~\cite{Gibbons:1985ac,Gibbons:1994vm,Gibbons:1994ff} and 
the evaporation process of the Hawking radiation~\cite{Koga:1995bs}.  
Even with such unusual behaviors, the uniqueness theorem of static black 
holes continues to be valid in this theory, viz, the spherically symmetric solution found by Gibbons and
Maeda~\cite{GM} exhausts all the asymptotically flat, static black hole
with a nondegenerate event horizon in the Einstein-Maxwell-dilaton 
theory~\cite{MasoodulAlam:1993ea,Gibbons:2002av}.

Despite the extensive work over the last two decades, a rotating black hole
solution in this theory has been yet available with the exception of
a slowly rotating approximate solution~\cite{Horne:1992zy} and a 
Kaluza-Klein black hole~\cite{Frolov:1987rj,Aliev:2008wv}. A widely used formalism for 
obtaining a new solution is the solution-generating method  
for the stationary spacetime, in which certain gravitational theories 
are dimensionally reduced to 3-dimensional gravity coupled to
scalar fields~\cite{Ernst:1967wx,Geroch:1972yt,Breitenlohner:1987dg}. 
In the Einstein-Maxwell theory, 
the target space of the harmonic maps is described by Bergmann metric
having the structure group isomorphic to coset 
${\rm SU(2,1)}/{\rm S}[{\rm U}(1, 1)\times {\rm U}(1)]$, which is large enough to contain the    
Ehlers-Harrison type transformations~\cite{Kinnersley:1977pg,Stephani:2003tm}. 
If an additional axisymmetry
is imposed the system becomes completely integrable, admitting a
variety of generation techniques~\cite{Geroch2,Belinsky:1971nt,Neugebauer:1979iw}. 
In the Einstein-Maxwell-dilaton theory, however,  
the target space is neither symmetric nor homogeneous (i.e., the coset
representation is impossible and the 
isometry group does not act transitively) for a generic
dilaton coupling~\cite{Galtsov:1995mb}. Furthermore, the additional 
axisymmetry fails to render the
system to be two-dimensionally integrable.  
This fact forbids us to get rotating black-hole solutions 
from simpler seed solutions following the  conventional procedure.  
In this paper, we adopt an alternative strategy by focusing on  
{\it supersymmetric} solutions.

Supersymmetric solutions in supergravity 
have performed an invaluable r\^ole in the progression of
non-perturbative r\'egime of string theory 
and the anti-de Sitter/conformal field theory correspondence. 
The supersymmetric solutions saturate  
the Bogomol'nyi-Prasad-Sommerfield (BPS) bound and 
are characterized by the existence of a super-covariantly constant
spinor referred to as a Killing spinor~\cite{Witten:1981mf,GH1982,PET}. 
One can identify the Killing spinor equations as the ``square root''
of field equations, so that supersymmetric solutions 
can be obtained relatively easily just by solving linear equations.  
As a matter of fact, we can systematically classify and sometimes can obtain  
{\it all} supersymmetric solutions.  
An initiated work is due to Tod, who 
inventoried all the BPS solutions admitting a nontrivial
Killing spinor in 4-dimensional $N=2$ supergravity~\cite{Tod}. 
Although reference~\cite{Tod} shed some light on the whole picture of
BPS solutions,  his method lacks utility in higher dimensions since the Newman-Penrose
formalism has been used therein. This difficulty can be overcome by the
seminal work of Gauntlett {\it et al.}~\cite{GGHPR}, in which general supersymmetric solutions in 5-dimensional
minimal supergravity were classified by making use of bilinears
constructed from a Killing spinor. Thereafter 
the classification program has achieved a remarkable development 
in diverse supergravities in various
dimensions~\cite{Gauntlett:2003fk,Gutowski:2003rg,Caldarelli:2003pb,GR,Gauntlett:2002fz,Bellorin:2006yr,Gran}.
This formulation has provided valuable tools for finding supersymmetric black holes~\cite{Gutowski:2004ez},
black rings~\cite{Elvang:2004rt}, and
for proving uniqueness theorem of certain black holes~\cite{Reall2002}.
It turns out that all the  supersymmetric black-hole solutions
have  universal properties such as strict stationarity and mechanical
equilibrium in the ungaged supergravities.  This means that black holes fail to
posses the trapped region 
(e.g., inside the Schwarzschild interior) and the ergoregion even if it
has a nonvanishing angular momentum.  
The mechanical equilibrium condition allows a multiple collection of
black holes, reflecting a ``no force'' situation between BPS objects~\cite{Hartle:1972ya}.   
The BPS configurations are thus very simple since  
supersymmetry prohibits any dynamical processes.

In this paper, we consider a simple model of Einstein-Maxwell-dilation
gravity  described by the action
\begin{align}
 S= \frac{1}{16\pi G }\int \D ^4 x \sqrt{-g }\left[R  -2 
(\nabla^\mu \phi)(\nabla_\mu \phi)- e^{-2\alpha \phi} F_{\mu\nu }
 F^{\mu\nu }  \right]\,, 
\label{action}
\end{align}
where $\phi$ is a dilaton field, $F=\D A$ is the Maxwell field and 
$\alpha $ controls the strength of the coupling of a dilaton to the
Maxwell field.    
The critical coupling  $\alpha=1$ arises 
by the truncation of $N=4$ supergravity~\cite{Gibbons:1982ih,GM,Garfinkle:1990qj}. 
Whereas,  the $\alpha=\sqrt 3$ case occurs via the Kaluza-Klein
compactification of 5-dimensional vacuum gravity.   
Nevertheless, it has been shown
that the theory admits a Bogomol'nyi-type inequality for an {\it arbitrary} coupling, 
and  allows a nontrivial ``Killing spinor'' of gravitino and dilatino 
when the inequality is saturated~\cite{GKLTT}. 
This fact strongly encourages us to speculate 
that the theory~(\ref{action}) can be embedded into some supergravity theories for
general coupling.

However, it has been known that the equilibrium solutions in~\cite{Gibbons:1982ih} do not
saturate this bound. This fact has given rise to some confusion in
the literature. 
In this paper we revisit the Bogomol'nyi bound 
and examine the integrability condition of the dilatino Killing spinor 
equation. A basic belief for the fermionic supersymmetry transformations is
that their integrability conditions guarantee the corresponding bosonic equations of motions.  
We argue that this consistency condition is satisfied only for
certain cases. Bering this remark in mind, we try to list all the supersymmetric vacua of this theory 
under the circumstances in which the consistency condition is
satisfied. This analysis unveils why the dyonic supersymmetric solutions 
with $\alpha =\sqrt 3$ are rotating~\cite{Tod2}. 
In the classification procedure we adopt a prescription of~\cite{GGHPR}, which is 
adequate also in the proof for the variant of positive energy theorem 
described below.  The supersymmetric differential relations explicitly  
show that the Arnowitt-Deser-Misner (ADM) mass coincides with the Komar integral associated with
a supersymmetric Killing vector. 
We shall also discuss physical 
properties of supersymmetric solutions.

The present paper is organized as as follows. 
In the next section, we give a brief overview on the
Einstein-Maxwell-dilaton gravity and discuss the Bogomol'nyi inequality. 
Section~\ref{sec:classification} is devoted to the systematic
construction of all BPS solutions, which fall into a timelike and a null
family. Some properties of BPS
solutions are analyzed in section~\ref{sec:pp}.
Section~\ref{sec:summary} concludes with several future prospects. 
A proof of an energy bound in arbitrary dimensions with a 
Kaluza-Klein coupling is given in appendix.

Throughout the paper, 
we use the mostly plus metric convention.  
Greek indices $\mu, \nu,...$ denote the spacetime 
indices, whereas Roman indices $a, b, ...$ 
refer to those in tangent space. 
The Hodge dual is denoted by star.
Gamma matrix $\gamma _\mu $ satisfies the Clifford algebra
$\{\gamma_\mu ,\gamma_\nu \}=2g_{\mu\nu }$. The 
antisymmetrized product 
is understood to be unit weight, e.g., 
$\gamma_{\mu\nu}=\gamma_{[\mu}\gamma_{\nu]}=(\gamma_\mu\gamma_\nu-\gamma_\nu\gamma_\mu)/2$
and so on. 
The chiral matrix is given by
$\gamma_5=-({\rm i}/4!)\epsilon_{abcd}\gamma^{abcd}$, so 
$\gamma_{\mu\nu\rho}=\I\epsilon_{\mu\nu\rho\sigma}\gamma^{\sigma}\gamma_5$
and 
$\gamma_{\mu\nu }=(\I/2)\epsilon_{\mu\nu\rho\sigma}\gamma^{\rho\sigma}\gamma_5 $ 
with $\epsilon_{0123}=1$.
We define the Dirac conjugate by 
$\bar \psi :={\rm i}\gamma^0 \psi^\dagger $.

\section{Energy bound in Einstein-Maxwell-dilaton gravity}

\subsection{Field equations}

The gravitational field equations derived from the action~(\ref{action}) are 
\begin{align}
R_{\mu\nu }-\frac 12 R g_{\mu \nu } &= T_{\mu\nu }  \nonumber\\
&= T_{\mu\nu }^{(\phi)} +T_{\mu\nu }^{(\rm em)} \,,
\label{Einsteineq}
\end{align}
where $T_{\mu\nu }$ is the total stress-energy tensor and 
\begin{align}
 T_{\mu\nu }^{(\phi)} &= 2 \left[ (\nabla_\mu \phi)( \nabla_\nu \phi)
 -\frac 12 g_{\mu\nu }(\nabla_\rho \phi )(\nabla^\rho \phi ) 
\right] \,, \\\
T_{\mu\nu }^{(\rm em)} &= 2 e^{-2\alpha\phi }
\left(F_{\mu \rho }{F_\nu }^\rho -\frac 14 g_{\mu\nu }F_{\rho\sigma
 }F^{\rho\sigma }\right) \,. 
\end{align}
The conservation equations for each stress-energy tensor lead to  
the Maxwell equations 
\begin{align}
 \nabla_\nu (e^{-2\alpha \phi }F^{\mu\nu })=0 \,,
\label{Maxwell}
\end{align}
and the dilaton evolution equation
\begin{align}
\nabla_\mu \nabla^\mu \phi +\frac \alpha 2 e^{-2\alpha \phi}F_{\mu\nu
 }F^{\mu\nu }=0\,.
\label{dilaton}
\end{align}
The action~(\ref{action}) is invariant under the discrete duality rotation,
\begin{align}
 \phi \to \tilde \phi = -\phi \,, \qquad F_{\mu\nu } \to
\tilde F_{\mu\nu } =e^{-2\alpha\phi }\star F_{\mu\nu 
 }\,.
\label{duality}
\end{align}
The continuous electric-magnetic duality symmetry in Einstein-Maxwell
theory is broken by the presence of a dilaton. 

It should be emphasized that 
the constant dilaton reduces not to the Einstein-Maxwell
system but to the Brans-Dicke-Maxwell theory with a Brans-Dicke constant
$\omega=-1$. The ordinary Einstein-Maxwell system is recovered when 
$\phi={\rm constant}$ and $F_{\mu\nu }F^{\mu\nu }=0$, or 
$\phi={\rm constant}$ and $\alpha=0$; otherwise the dilaton field
equation~(\ref{dilaton}) is not satisfied. 
For $\alpha =1$, the action~(\ref{action}) corresponds to the  
truncated action of $N=4$ supergravity. The action for
$\alpha=\sqrt 3$ is the Kaluza-Klein reduction of five-dimensional
vacuum gravity. When $\alpha=\sqrt{p/(p+2)}$ ($p=0,1,2,...$) 
the action arises from the $\mathbb T^p$ compactification of
(static truncation of) $(4+p)$-dimensional Einstein-Maxwell theory~\cite{Gibbons:1994vm}, i.e.,
the dimensional reduction along the $p$-brane metric.

\subsection{BPS inequality}

At least for the aforementioned values of $\alpha$,
there are underlying supergravity theories. Still,  
the Einstein-Maxwell-dilaton gravity~(\ref{action}) enjoys a
Bogomol'nyi-type inequality for general values of $\alpha $, 
as shown by Gibbons {\it et al.}~\cite{GKLTT}. 
We begin by a brief review about their argument and move to the detailed
discussion about the BPS inequality.
  
Following the standard prescription of the positive energy
theorem~\cite{Witten:1981mf,GH1982,PET}, define a Nester-like
anti-symmetric tensor in terms of a  super-covariant derivative 
$\hat \nabla_\mu $ acting on a (commuting) spinor $\epsilon $ by
\begin{align}
 \hat E^{\mu\nu }=- {\rm i} \left(\bar \epsilon \gamma^{\mu\nu\rho }\hat
 \nabla_\rho \epsilon - 
\overline{\hat\nabla_\rho \epsilon }
\gamma^{\mu\nu\rho }\epsilon  \right) \,. 
\end{align} 
Here, the operator $\hat \nabla_\mu $ is defined by 
\begin{align}
 \hat \nabla_\mu \epsilon = 
 \left(\nabla_\mu +\frac{\rm i}{4\sqrt{1+\alpha ^2 }}e^{-\alpha \phi}
\gamma ^{ab}\gamma_\mu F_{ab}\right)\epsilon \,,
\end{align}
which specifies the ``variation of gravitino.'' When acting on a spinor,  
the covariant derivative $\nabla_\mu $ is given in terms of a torsion-free spin
connection $\omega_{\mu ab}$ as 
\begin{align}
 \nabla_\mu \epsilon =\left(\partial_\mu +\frac 14
 {\omega_\mu}^{ab}\gamma_{ab }\right)\epsilon \,,
\end{align} 
which obeys the Leibniz rule 
\begin{align}
\nabla_\mu (\bar \epsilon _1 \gamma_{\mu_1}\cdots \gamma_{\mu
 _n}\epsilon _2 )
&= 
\overline{\nabla_\mu \epsilon_1 }
\gamma_{\mu_1}\cdots
 \gamma_{\mu_n}\epsilon_2 +\bar \epsilon_1  \gamma_{\mu_1}\cdots
 \gamma_{\mu_n}  \nabla_\mu\epsilon _2 \,, \nonumber \\
\nabla_\mu (
\bar \epsilon _1 \gamma_5 \gamma_{\mu_1}\cdots \gamma_{\mu
 _n}\epsilon _2 )
&=
\overline{\nabla_\mu \epsilon_1} \gamma_5 \gamma_{\mu_1}\cdots
 \gamma_{\mu_n}\epsilon _2 +\bar \epsilon _1  \gamma_5\gamma_{\mu_1}\cdots
 \gamma_{\mu_n}   \nabla_\mu\epsilon _2 \,.
\label{Liebniz}
\end{align}
Observe that $\hat E^{\mu\nu }$ decompose as 
\begin{align}
 \hat E^{\mu\nu }=E^{\mu\nu }+H^{\mu\nu }\,,
\end{align}
where 
$
E^{\mu\nu }=-{\rm i}(\bar\epsilon\gamma^{\mu\nu\rho}\nabla_\rho
\epsilon -\overline{\nabla_\rho\epsilon }\gamma^{\mu\nu\rho}\epsilon)
$
is an ordinary Nester 2-tensor and $H^{\mu\nu }$ represents the
electromagnetic contribution,  
\begin{align}
 H^{\mu\nu } &= -\frac{2e^{-\alpha \phi }}{\sqrt{1+\alpha^2 }}
\left(\bar\epsilon\epsilon F^{\mu\nu } -{\rm i}\bar \epsilon
 \gamma_5\epsilon \star F^{\mu\nu } \right) \,. 
\label{em_H}
\end{align}

Reference~\cite{GKLTT} also introduced the ``variation of dilatino''
by\footnote{
Note that 
the conventions of the present paper differs from~\cite{GKLTT}, where 
the gamma matrix and the Dirac conjugate are defined by 
$\{\gamma_\mu ,\gamma_\nu \}=-2g_{\mu\nu }$ and 
$\bar \psi =\psi ^\dagger \gamma^0 $. 
Equation~(\ref{dilatino_var}) also corrects the typo 
in~\cite{GKLTT}. }
\begin{align}
 \delta \lambda := \frac{1}{\sqrt 2} 
\left(\gamma ^\mu \nabla_\mu \phi -\frac{{\rm i} \alpha }{2\sqrt{1+\alpha ^2}}
e^{-\alpha \phi }\gamma^{ab}F_{ab} \right)\epsilon \,. 
\label{dilatino_var}
\end{align}
Here, the specific factors appearing in~(\ref{KS})
and (\ref{dilatino}) have been chosen a posteriori in order
to give an energy bound.

Consider an asymptotically flat spacetime to which 
an ADM
4-momentum can be assigned~\cite{Witten:1981mf,ADM}. 
Choose a spatial hypersurface $\Sigma $ with a 
future-pointing unit normal $n^\mu $ and let 
 $\partial \Sigma $ be its boundary at spatial infinity.\footnote{
Although we have supposed that $\Sigma $ has no interior boundary
corresponding to the black hole horizon, this condition can be
relaxed~\cite{GKLTT} (see also~\cite{Gibbons:1982jg}). }  
Assume that $\epsilon $ asymptotes to a constant spinor
$\epsilon_\infty $ and that the dilaton falls off
to zero at spatial infinity.  Using Stokes' theorem, 
it is found that 
\begin{align}
-\int _\Sigma \D \Sigma n_\mu \nabla_\nu \hat E^{\mu \nu } &= 
\frac 12 \int_{\partial \Sigma }\D S_{\mu\nu }\hat E^{\mu\nu }
 \nonumber \\
&= \frac 12 
\int _{\partial\Sigma }\D S_{\mu\nu } E^{\mu\nu }
 -\frac 1{\sqrt{1+\alpha^2 }}\int _{\partial\Sigma } \D
 S_{\mu\nu } \left( \bar \epsilon_\infty \epsilon_\infty F^{\mu\nu }
-{\rm i}\bar \epsilon_\infty \gamma_5\epsilon_\infty \star F^{\mu\nu }
 \right)\nonumber \\ 
&=-{\rm i}\bar \epsilon_\infty \gamma^\mu \epsilon_\infty P_\mu
 -\frac{1}{\sqrt{1+\alpha^2 }}\bar \epsilon _\infty (Q_e-{\rm i}\gamma_5 Q_m
 )\epsilon_\infty \,,\label{Gausseq}
\end{align}
where $\D S_{\mu\nu }$ is the element of 2-sphere at infinity.
$P_\mu $ denotes the ADM
 4-momentum~\cite{Witten:1981mf,ADM} and 
\begin{align}
Q_e = \int_{\partial\Sigma } \D S_{\mu\nu }F^{\mu\nu } \,,
 \qquad 
Q_m = \int_{\partial\Sigma } \D S_{\mu\nu }\star  F^{\mu\nu } \,, 
\end{align}
are the total electric and magnetic charges, respectively. 
A straightforward but rather tedious calculation shows
that
\begin{align}
 \nabla_\nu \hat E^{\mu\nu }=&
2{\rm i}\overline{\hat\nabla_\rho \epsilon}
 \gamma^{\mu\nu\rho }\hat \nabla_\nu
 \epsilon + 2{\rm i}\overline{\delta\lambda }\gamma^\mu \delta \lambda 
-\left({R^\mu }_\nu -\frac 12 R{\delta^\mu }_\nu  -{T^\mu }_\nu \right) 
({\rm i} \bar \epsilon \gamma^\nu \epsilon )
\nonumber \\
& 
-\frac{2}{\sqrt{1+\alpha^2 }}\left[e^{\alpha\phi }\nabla_\nu (e^{-2\alpha\phi }F^{\mu\nu }) \bar
 \epsilon \epsilon -e^{-\alpha \phi}(\nabla_\nu \star F^{\mu \nu })
({\rm i}\bar \epsilon \gamma_5 \epsilon )\right] \,.
\end{align}
Relations~(\ref{useful_rel1})--(\ref{useful_rel8}) in the next section  
are of great help to derive this equation. 
The last three terms will vanish provided Einstein's equations, the Maxwell
equations and the Bianchi identity are imposed. 
Then the volume integral of the left-hand side of (\ref{Gausseq})
can be written as a sum of non-negative terms for $\epsilon $ satisfying
the (modified) Dirac-Witten equation 
\begin{align}
 \gamma^{a} \hat D_{a} \epsilon =0 \,, 
\end{align}
where $\hat D$ is the projection of super-covariant derivative
$\hat \nabla $ onto $\Sigma $. 
It follows that 
the right hand side of~(\ref{Gausseq}) has to have non-negative
eigenvalues, giving rise to a suggestive inequality
\begin{align}
M\ge \frac{1}{\sqrt{1+\alpha^2 }}\sqrt{Q_e^2+Q_m^2 }\equiv M_{\rm BPS}\,,  
\label{BPS}
\end{align}
where $M=\sqrt{-P_\mu P^\mu }$ is the ADM mass. 
In the context of supergravity, $Q_e$ and $Q_m$ should enter the algebra of 
global  supersymmetry transformations as central charges. 
The above lower bound is attained if and only if 
there exists a nontrivial spinor $\epsilon $ satisfying
the gravitino Killing spinor equation
\begin{align}
 \left(\nabla _\mu +\frac{\rm i}{4\sqrt{1+\alpha ^2 }}e^{-\alpha \phi}
\gamma ^{ab}\gamma_\mu F_{ab}\right)\epsilon =0 \,,
\label{KS}
\end{align}
and the dilatino  Killing spinor equation
\begin{align}
 \left(\gamma ^\mu \nabla_\mu \phi -\frac{{\rm i} \alpha }{2\sqrt{1+\alpha ^2}}
e^{-\alpha \phi }\gamma^{ab}F_{ab} \right)\epsilon =0 \,. 
\label{dilatino}
\end{align}
These can be viewed as supersymmetry transformations which leave the 
bosonic background invariant.

The resulting Killing spinor equations and the energy bound~(\ref{BPS}) strongly imply that the 
theory~(\ref{action}) might be embedded into some supergravity theory
for the general value of $\alpha $~\cite{PET}. 
We are now going to claim, however, that this might be too optimistic an estimate.
To illustrate, let us consider the multiple black hole solution found
in~\cite{Gibbons:1982ih}, 
\begin{align}
\D s^2=- H_1^{-1}H_2^{-1}\D t^2+H_1H_2 \D \vec x^2 \,, 
\label{Gibbons}
\end{align} 
where $H_1$ and $H_2$ are arbitrary harmonics on the 
Euclid 3-space  $\mathbb R^3$, and
\begin{align}
 A=\frac{1}{\sqrt 2} \left(\frac{\D t}{H_1}+\vec A \cdot \D \vec x\right) \,, 
 \qquad \vec \nabla \times \vec A = \vec \nabla H_2 \,, \qquad 
\phi =-\frac{1}{2} \ln\left(\frac{H_1}{H_2}\right) \,.
\end{align}
Here and hereafter,  the 3-dimensional vector notation will be used
for quantities of 3-dimensional Euclid space. 
The metric~(\ref{Gibbons}) solves the field equations~(\ref{Einsteineq}),
(\ref{Maxwell}) and (\ref{dilaton}) with $\alpha=1$, which is the 
distinguished value predicted by string theory.   
Two functions $H_1$ and $H_2$ obey Laplace equations, so the 
feature of force balance is appropriately captured. At first sight, it
therefore seems reasonable to expect that  this solution would 
saturate the bound~(\ref{BPS}). 
Contrary to our intuition, this is not the case~\cite{Rasheed:1995zv}. 
Consider multiple point sources 
\begin{align}
H_1=1+ \sum_k \frac{\sqrt 2 Q_e^{(k)}}{|\vec x-\vec x_{(k)}|} \,, 
\qquad 
H_2 = 1+ \sum_k \frac{\sqrt 2 Q_m^{(k)}}{|\vec x-\vec x_{(k)}'|} \,, 
\end{align}
where  $\vec x_{(k)}$ and 
$\vec x_{(k)}'$ represent the loci of point sources.  
One immediately finds that the metric is asymptotically flat,
 $Q_e=\sum_kQ_e^{(k)}$ and $Q_m=\sum_kQ_m^{(k)}$ correspond to the 
total electric and magnetic charges, in terms of which the ADM mass is
given by $M=(Q_e+Q_m)/\sqrt 2$. This is strictly above the
lower bound~(\ref{BPS}).\footnote{
If the ``scalar charge'' $\Sigma $ is introduced by the asymptotic value
of the scalar field as $\phi \sim \pm \Sigma/|\vec x|$, 
the nonextremal metric in~\cite{Gibbons:1982ih} admits an inequality
$M^2 +\Sigma^2 \ge Q_e^2+Q_m^2$, which is saturated by the BPS
state~(\ref{Gibbons}). It is worthwhile to emphasize that this inequality differs from~(\ref{BPS}) in
philosophy: the Bogomol'nyi inequality~(\ref{BPS}) is expressed only in
terms of global charges, while the above force-balance condition
involves the scalar charge which is inherently secondary since it is not
defined covariantly by the two-sphere surface integral at infinity. 
}
The metric~(\ref{Gibbons}) has provided potential confusions in the
literature. In reference~\cite{Rasheed:1995zv}, a different 
expression of the Bogomol'nyi-type bound 
\begin{align}
M= \frac{1}{\sqrt{1+\alpha^2 }}\left[Q_e^n+Q_m^n\right]^{1/n}\,, \qquad n=\frac{2}{1+\log_2
 (1+\alpha^2 )} \,,
\end{align}
is conjectured  from the 
force-balance point of view.

This puzzling issue is best understood as follows.
Acting $\gamma^\nu \nabla_\nu $ to~(\ref{dilatino}) and
using~(\ref{KS}), 
we obtain  
\begin{align}
&\left[\nabla_\mu \nabla^\mu \phi +\frac{\alpha}2
e^{-2\alpha\phi }F_{\mu\nu}F^{\mu\nu }
+\frac{{\rm i}\gamma_5  
\alpha (\alpha^2 -3)}{2 (1+\alpha^2 )}
F_{\mu\nu }\star F^{\mu\nu }
\right.\nonumber \\
&\left. ~~  
 -\frac{{\rm i}\alpha e^{-\alpha\phi }}
{2\sqrt{1+\alpha^2 }} \left\{\gamma^{\mu\nu\rho }\nabla_{[\mu
 }F_{\nu\rho] }-2e^{2\alpha\phi }\gamma_\mu \nabla_\nu (e^{-2\alpha\phi
 }F^{\mu\nu })\right\}
\right]\epsilon =0 \,.
\label{integrability_dil}
\end{align}  
Accordingly, even if the Bianchi identity $\D F=0 $ and the Maxwell equations
$\D \star (e^{-2\alpha\phi }F)=0$ are satisfied, 
the integrability condition of the dilatino
equation~(\ref{integrability_dil}) does {\it not} guarantee the dilaton
equations of motion~(\ref{dilaton}) apart from $\alpha =0, \sqrt 3$ or  
$F_{\mu\nu }\star F^{\mu\nu }=0$. In this sense,  
{\it the dilatino equation is not the proper 
``square root'' of the dilaton field equation}. 
The dyonic solution~(\ref{Gibbons}) is not supersymmetric  
in spite of string motivated case $\alpha=1$
since it does not satisfy $F_{\mu\nu }\star F^{\mu\nu }= 0$.

A major cause of this apparent variance may be attributed to the absence of the 
axion field in the theory. The effective theory of heterotic string
indeed involves the axion field, which  couples to $F_{\mu\nu }\star F^{\mu\nu }$ term 
in the Lagrangian. It therefore cannot be consistently truncated  
unless  $F_{\mu\nu }\star F^{\mu\nu}=0$~\cite{Rasheed:1995zv,Galtsov:1994pd} 
(see~\cite{Rogatko:1995bz} for a proof of the Bogomol'nyi inequality in the
Einstein-Maxwell-dilaton-axion system). 
This observation leads to speculate that the Gibbons solution~(\ref{Gibbons}) is the BPS
solution to some truncation of different supergravity theory, 
rather than the truncation of Einstein-Maxwell-dilaton-axion gravity.  
It seems interesting to examine which supergravity has~(\ref{Gibbons})
as a BPS solution. But addressing this issue  
is beyond the scope of the present paper.

Nonetheless, the configurations which saturate the bound~(\ref{BPS}) can be identified as 
ground states that minimize the energy for fixed charges, irrespective of whether the
Einstein-Maxwell-dilaton theory~(\ref{action}) has a supergravity origin
or not.  Tod has shown~\cite{Tod2} that the supersymmetric solutions 
with $\alpha\ne \sqrt 3$ are necessarily static and  
described by the Gibbons-Maeda solution~\cite{GM}. 
But the analysis~\cite{Tod2} has unsettled as to why the rotating solution is
not allowed for $\alpha\ne \sqrt 3$. In the next section,, 
we classify solutions admitting a Killing spinor  
which satisfies the 1st-order differential equations~(\ref{KS}) and
(\ref{dilatino}) via the modern analysis initiated in~\cite{GGHPR}.

\section{Supersymmetric solutions}
\label{sec:classification}

In this section, we shall classify the supersymmetric solutions 
following the work of Caldarelli and Klemm~\cite{Caldarelli:2003pb}.
A basic strategy for the classification of BPS solutions is to assume
the existence of at least one Killing spinor and construct its bilinear
tensor quantities. These satisfy a number of algebraic and differential 
conditions, which can be used to deduce the bosonic constituents. 
The final result in this section coincides with the work of
Tod~\cite{Tod2} utilizing the Newman-Penrose technique, 
so the readers interested in the physical properties of
BPS solutions may skip this section (though discussion in the next
section requires techniques and equations derived in this section). 
We point out explicitly that the dyonic condition is equivalent to the 
failure of the stationary Killing field being hypersurface-orthogonal.
This issue has not been argued in the reference~\cite{Tod}.

\subsection{Differential forms constructed from a Killing spinor}

Given a commuting spinor $\epsilon $, we can define the following 
bilinear bosonic differential forms~\cite{Caldarelli:2003pb}
\begin{align}
 \textit{a scalar} & ~~~ E:= \bar \epsilon \epsilon \,,
\label{DF1} \\
 \textit{a pseudo scalar}  &~~~ B:=  {\rm i}\bar \epsilon\gamma_5
 \epsilon \,, \label{DF2}\\
 \textit{a vector}  &~~~ V_\mu :=  {\rm i}\bar \epsilon\gamma_\mu
 \epsilon \,, \label{DF3}\\
 \textit{a pseudo vector}  &~~~ a_\mu :=  {\rm i}\bar
 \epsilon\gamma_5\gamma_\mu 
 \epsilon \,, \label{DF4}\\
 \textit{an anti-symmetric tensor}  &~~~ \Phi_{\mu\nu}:=  {\rm i}\bar
 \epsilon\gamma_{\mu\nu }
 \epsilon \,, \label{DF5}
\end{align}
Here we have introduced the factor ``${\rm i}$'' to ensure these differential forms to be
real in our convention. Since 
$\{{\bf 1},\gamma_5, \gamma_\mu ,\gamma_\mu \gamma_5, \gamma_{\mu \nu }\}$
span the basis of Clifford algebra, any other differential forms can be 
built from linear combination of above quantities. 

A (Dirac) spinor $\epsilon $ has a real dimension 8, whereas 
$(E, B, V_\mu ,a_\mu , \Phi_{\mu\nu })$ sum up to have 16 components. 
This means that these bilinears are not all independent. In fact, 
viewing $\epsilon \bar \epsilon $ as a $4\times 4$ matrix, 
it can be expanded by gamma-matrix basis as 
\begin{align}
 4 \epsilon \bar \epsilon =E {\bf 1}-{\rm i}V^\mu \gamma_\mu +
\frac{\rm i}2\Phi^{\mu\nu }\gamma_{\mu\nu }+{\rm i}a^\mu
 \gamma_5\gamma_\mu -{\rm i}B \gamma_5 \,,
\label{Fierz}
\end{align}
which implies
\begin{align}
{\rm i}V^\mu \gamma_\mu \epsilon &=-{\rm i}a^\mu \gamma_5\gamma_\mu
 \epsilon =-(E+{\rm i }B \gamma_5 )\epsilon \,,  \qquad 
{\rm i}\Phi^{\mu \nu }\gamma_{\mu \nu} 
\epsilon  =2(E-{\rm i}B\gamma_5 )\epsilon \,. 
\label{projection}
\end{align}
Contraction with $\bar \epsilon $ gives 
\begin{align}
& f:=-V^\mu V_\mu =a^\mu a_\mu =E^2+B^2\,,\label{lapse}\\ 
&E^2-B^2 =\tfrac 12 \Phi_{\mu \nu }\Phi^{\mu \nu }\,. 
\label{Phisq}
\end{align}
We can find that $V^\mu $ is everywhere causal,
while $a^\mu $ cannot be timelike.
The possibility of $V^\mu \equiv 0 $ can be eliminated by noticing 
$V^0 =\epsilon^\dagger \epsilon >0$ for a nonvanishing Killing spinor. 
This also signifies that $V^\mu $ is future-directed. 
Contracting $\bar \epsilon \gamma_5 $ to (\ref{projection}), 
it is shown that  $V$ and $a$ are orthogonal $V^\mu a_\mu =0$.

Using~(\ref{Fierz}) and 
availing ourselves of the useful expressions,
the differential forms (\ref{DF1})--(\ref{DF5}) 
constructed from a commuting spinor $\epsilon $ satisfy 
\begin{subequations}
\label{useful_rel}
\begin{align}
\label{useful_rel1}
 \bar \epsilon \gamma_\mu \gamma_\nu \epsilon & 
=Eg_{\mu \nu }-{\rm i}\Phi_{\mu \nu} \,, \\
 \bar \epsilon \gamma_5 \gamma_\mu \gamma_\nu \epsilon &= -{\rm i}B g_{\mu \nu
 }+\star \Phi_{\mu \nu }\,, \\
\bar \epsilon \gamma_{\mu \nu }\gamma_\rho \epsilon
& =-\epsilon_{\mu\nu\rho\sigma }a^\sigma -2{\rm i}V_{[\mu }g_{\nu
 ]\rho}\,, \\
 \bar \epsilon \gamma_5 \gamma_{\mu \nu }\gamma_\rho \epsilon
& =-\epsilon_{\mu\nu\rho\sigma }V^\sigma -2{\rm i}a_{[\mu }g_{\nu
 ]\rho}\,, \\
\bar \epsilon \gamma_{\mu \nu }\gamma_{\rho\sigma }\epsilon &= 
-B\epsilon_{\mu\nu \rho\sigma }+2{\rm i}(\Phi_{\mu [\rho  }g_{\sigma
 ]\nu}- g_{\mu [\rho  }\Phi_{\sigma ]\nu})-2 E g_{\mu [\rho }g_{\sigma ]
 \nu}\,, \\
\bar \epsilon \gamma_5 \gamma_{\mu \nu }\gamma_{\rho\sigma }\epsilon &= 
-{\rm i}E\epsilon_{\mu\nu \rho\sigma }+
2\epsilon _{\mu\nu\lambda [\rho}{\Phi^\lambda }_{\sigma] }
+2{\rm i} B g_{\mu [\rho }g_{\sigma ]
 \nu}\,, \\
\bar \epsilon \gamma_\mu \gamma_{\rho\sigma }\gamma_\nu \epsilon &= 
-B\epsilon_{\mu\nu \rho \sigma }-2{\rm i}\Phi_{\mu[\rho }g_{\sigma]\nu}
-2{\rm i}g_{\mu[\rho }\Phi_{\sigma]\nu}-{\rm i}g_{\mu\nu}\Phi_{\rho
 \sigma }+2Eg_{\mu[\rho }g_{\sigma] \nu }\,,\\
\bar \epsilon \gamma_5 \gamma_\mu \gamma_{\rho\sigma }\gamma_\nu
 \epsilon &= 
-{\rm i}E\epsilon_{\mu\nu \rho\sigma }-2{\rm i}Bg_{\mu [\rho }g_{\sigma
 ]\nu}+2\epsilon _{\mu\nu\lambda [\rho}{\Phi^\lambda }_{\sigma] }
+ g_{\mu\nu}\star \Phi_{\rho\sigma}\,,
\label{useful_rel8}
\end{align}
\end{subequations}
it is straightforward to derive the following algebraic constraints 
\begin{subequations}
\begin{align}
 EV_\mu &=\star \Phi_{\mu\nu }a^\nu \,, \qquad  \qquad \quad ~
 Ea_\mu =\star \Phi_{\mu\nu }V^\nu \,,
\label{alg_cons_1}\\
 BV_\mu &= \Phi_{\mu\nu }a^\nu 
\,, \qquad \qquad \qquad Ba_\mu =\Phi_{\mu\nu}V^\nu \,,\label{alg_cons_2}\\
EB &=-\tfrac{1}{4}\Phi_{\mu\nu }\star  \Phi^{\mu\nu }\,, \label{alg_cons_3}\\
E\Phi_{\mu\nu }&=-\epsilon_{\mu \nu \rho \sigma }V^\rho a^\sigma 
+B\star \Phi_{\mu\nu }\,,\label{alg_cons_4}\\
{\Phi_{(\mu }}^\rho \star \Phi_{\nu)\rho }
&=\tfrac 14 g_{\mu\nu }\Phi_{\rho\sigma 
 }\star \Phi^{\rho\sigma} \,.\label{alg_cons_5}
\end{align}
\end{subequations}
Upon using~(\ref{lapse}) and (\ref{Phisq}), 
a little bit amount of calculation shows 
\begin{align}
 \Phi_{\mu\rho }{\Phi_\nu }^\rho &=V_\mu V_\nu -a_\mu a_\nu +g_{\mu\nu
 }E^2 \,, \qquad 
\star  \Phi_{\mu\rho }\star {\Phi_\nu }^\rho =V_\mu V_\nu -a_\mu a_\nu +g_{\mu\nu
 }B^2 \,.
\label{PhiPhi}
\end{align}
This relation will be of use 
for the classification of null class.

Let us turn to the analysis of differential relations. 
Now suppose that $\epsilon $ satisfies the Killing spinor
equation~(\ref{KS}). 
Noticing~(\ref{Liebniz}), we can derive the following differential constraints
\begin{subequations}
\begin{align}
 \nabla_\mu E &= \frac{e^{-\alpha \phi}}{\sqrt{1+\alpha ^2 }} F_{\mu \nu }V^\nu \,, 
\label{derv_E}\\
 \nabla_\mu B &=-\frac{e^{-\alpha \phi}}{\sqrt{1+\alpha ^2 }} 
\star  F_{\mu\nu }V^\nu \,, 
\label{derv_B}\\
\nabla_\mu V_\nu &=\frac{e^{-\alpha \phi}}{\sqrt{1+\alpha ^2 }}
\left(- EF_{\mu \nu }+B\star F_{\mu\nu } \right) \,,
\label{derv_V}\\
\nabla_\mu a_\nu &=\frac{e^{-\alpha \phi}}{\sqrt{1+\alpha ^2 }}
\left(-2 {F_{(\mu }}^\rho \star \Phi_{\nu)\rho }
+\frac 12 g_{\mu\nu }F_{\rho\sigma }\star \Phi^{\rho\sigma }
\right) \,,
\label{derv_a}\\
\nabla_\mu \Phi_{\nu\rho }&=-\frac{e^{-\alpha \phi}}{\sqrt{1+\alpha
 ^2 }}
\left( a_\mu \star F_{\nu\rho } +2\epsilon_{\nu\rho \sigma [\mu
 }{F_{\tau ] }}^\sigma a^\tau 
\right)
 \,.  
\label{derv_Phi}
\end{align}
\end{subequations}
We can thus identify $E$ and $B$ as the electric and magnetic potentials,
respectively. Equation~(\ref{derv_V}) indicates that 
$V^\mu $ is a Killing vector 
\begin{align}
\nabla_{(\mu }V_{\nu)}=0 \,.
\end{align}
From~(\ref{derv_a}) we  find 
$\nabla_{[\mu }a_{\nu ]}=0$, i.e.,  $a_\mu $ is a pure gradient vector.

Next, let us look into the dilatino equation~(\ref{dilatino}). 
Contracting it with $\bar \epsilon $, $\bar \epsilon \gamma_5$, 
$\bar \epsilon \gamma_\mu $, $\bar \epsilon \gamma_5\gamma_\mu $ and 
$\bar \epsilon \gamma_{\mu \nu }$ 
we obtain the following relations 
\begin{subequations}
\label{dil_con}
\begin{align}
 V^\mu \nabla _\mu \phi &=0\,, \label{dil_con_1} \\
\alpha \Phi_{\mu \nu }F^{\mu \nu } & =0 \,,  
\label{dil_con_12}\\
a^\mu \nabla_\mu \phi +\frac{\alpha e^{-\alpha \phi}}{2\sqrt{1+\alpha ^2}}
F_{\mu \nu }\star  \Phi^{\mu\nu }&=0 \,,
\label{dil_con_2}\\
E \nabla _\mu \phi -\frac{\alpha e^{-\alpha \phi}}{\sqrt{1+\alpha
 ^2}}F_{\mu\nu }V^\nu &=0 \,, \label{dil_con_31} \\ 
 \Phi_{\mu \nu }\nabla^\nu \phi - 
\frac{\alpha e^{-\alpha \phi}}{\sqrt{1+\alpha ^2}}
\star F_{\mu\nu }a^\nu &= 0 \,, 
\label{dil_con_32}\\
B\nabla_\mu \phi - \frac{\alpha e^{-\alpha
 \phi}}{\sqrt{1+\alpha ^2}}\star F_{\mu\nu }V^\nu &=0 \,,
 \label{dil_con_41} \\
\star  \Phi_{\mu\nu }\nabla^\nu \phi - \frac{\alpha e^{-\alpha
 \phi}}{\sqrt{1+\alpha ^2}}F_{\mu\nu}a^\nu& = 0  \,, 
\label{dil_con_42}\\
2V_{[\mu }\nabla_{\nu ]}\phi -\frac{\alpha
 e^{-\alpha\phi}}{\sqrt{1+\alpha ^2}}
\left(B \star F_{\mu\nu}+E F_{\mu \nu }
 \right) &=0 \,, \label{dil_con_51} \\
\epsilon_{\mu \nu\rho\sigma }a^\rho \nabla^\sigma \phi  
+\frac{2\alpha e^{-\alpha\phi}}{\sqrt{1+\alpha ^2}}
{F_{[\mu }}^\rho \Phi_{\nu ]\rho } &=0 \,.
\label{dil_con_52}
\end{align}
\end{subequations}
Contraction with $\bar \epsilon\gamma_5\gamma_{\mu \nu }$ 
yields the duals of (\ref{dil_con_51}) and (\ref{dil_con_52}).
When the Bianchi identity $\D F=0$ and the Maxwell equations
$\D \star (e^{-2\alpha\phi }F)=0$ are satisfied, 
equations~(\ref{derv_E}), (\ref{derv_B}), (\ref{dil_con_1}), 
(\ref{dil_con_31}) and (\ref{dil_con_41}) give 
\begin{align}
 \mas L_V F =0 \,, 
\qquad
 \mas L_V\star F=0
 \,,\qquad 
 \mas L_V \phi =0 \,, 
\end{align}
where  
$\mas L_V=\D i_V+i_V \D $ is the Lie derivative acting on form fields and 
$i_V$ is the interior product. It turns out that 
a vector field $V$ constructed from a Killing spinor 
generates the symmetry of all the bosonic constituents 
$(g_{\mu\nu }, F_{\mu\nu }, \phi)$. 
This is not an obvious result since the Killing symmetry just requires 
that $\mas L_V F$ is proportional to $\star F$ even in the 
Einstein-Maxwell system~(see Theorem 11.1 in~\cite{Stephani:2003tm}).

To proceed, we will examine separately the cases where 
the Killing vector is  timelike or null. 
The algebraic and differential constraints derived in this section are
solved for each case.

\subsection{Timelike family}
\label{sec:timelike}

Let us begin by the analysis for the case of timelike $V$, i.e., $f$ is
nowhere nonvanishing which we take $f>0$.  
Equations~(\ref{alg_cons_1}) and (\ref{alg_cons_2}) can
be solvable for $\Phi_{\mu \nu }$, giving  
\begin{align}
 \Phi_{\mu\nu  }=\frac{1}{f}\left(2BV_{[\mu  }a_{\nu]
 }-E\epsilon_{\mu\nu \rho\sigma }V^\rho a^\sigma \right)\,, \qquad
\star  \Phi_{\mu\nu  }=\frac{1}{f}\left(2EV_{[\mu }a_{\nu]
 }+B\epsilon_{\mu\nu \rho\sigma }V^\rho a^\sigma \right)\,.
\label{Phi_timelike}
\end{align}
These expressions are consistent with other
equations~(\ref{Phisq}), (\ref{alg_cons_3})--(\ref{alg_cons_5}) and
(\ref{PhiPhi}).  
Analogously equations (\ref{derv_E}) and (\ref{derv_B}) combine to give
\begin{align}
 F_{\mu \nu }=\frac{e^{\alpha \phi}\sqrt{1+\alpha ^2}}{f}\left(2V_{[\mu }\nabla_{\nu]
 }E+\epsilon_{\mu\nu\rho\sigma }V^\rho \nabla^\sigma  B \right)\,, \qquad
\star  F_{\mu\nu }=\frac{e^{\alpha \phi}\sqrt{1+\alpha ^2}}{f}\left(-2V_{[\mu }\nabla_{\nu]
 }B+\epsilon_{\mu\nu\rho\sigma }V^\rho \nabla^\sigma  E \right)\,.
\label{F_timelike}
\end{align}
From these expressions, one can easily verify
\begin{align}
F^{\mu\nu }F_{\mu\nu }&=\frac{2e^{2\alpha \phi
 }(1+\alpha^2)}{f}\left[(\nabla B)^2 -(\nabla E)^2\right]\,, &
F_{\mu\nu }\star F^{\mu\nu }&= \frac{4e^{2\alpha \phi
 }(1+\alpha^2)}{f}\nabla_\mu E\nabla^\mu B \,, \nonumber \\ 
F_{\mu\nu }\star \Phi^{\mu\nu }&=\frac{\sqrt{1+\alpha^2}e^{\alpha\phi
 }}{f}a^\mu\nabla_\mu (B^2-E^2)\,, 
&
 \Phi_{\mu\nu }F^{\mu\nu
 }&=-\frac{2\sqrt{1+\alpha^2}e^{\alpha\phi }}{f}a^\mu \nabla_\mu (EB)
 \,.
\end{align}

Substituting~(\ref{Phi_timelike}) and (\ref{F_timelike}) into 
(\ref{derv_V}) and (\ref{derv_a}), we obtain
\begin{align}
 \nabla_\mu V_\nu =& f^{-1} \left[-V_{[\mu }\nabla_{\nu]} f
-\epsilon_{\mu\nu\rho\sigma }V^\rho (E\nabla^\sigma B-B\nabla^\sigma E
 )\right]\,, \label{dV_timelike}\\
\nabla_\mu a_\nu =& -\tfrac 12 g_{\mu\nu }a^\rho \nabla_\rho (\ln f)
+a_{(\mu }\nabla_{\nu)} \ln f-f^{-2}V_\mu V_\nu a^\rho \nabla_\rho f 
+2 f^{-2}V_{(\mu }\epsilon_{\nu )\rho\sigma\tau }
(E\nabla ^\rho B-B\nabla^\rho E)V^\sigma a^\tau \,.
\label{da_timelike}
\end{align}
Using (\ref{Phi_timelike}), (\ref{F_timelike}),
(\ref{dV_timelike}) and (\ref{da_timelike}), 
a lengthy calculation shows that (\ref{derv_Phi}) is fulfilled automatically.

Inserting (\ref{F_timelike}) into 
the Maxwell equation $\D \star (e^{-2\alpha \phi }F)=0$ and the Bianchi
identity $\D F=0$, we find  
\begin{align}
f^2 \nabla ^\mu (f^{-1}\nabla_\mu E)+ \Omega_\mu \nabla^\mu
 B-\alpha f\nabla_\mu \phi \nabla^\mu E &=0 \,, 
\label{Maxeq_decomp}\\
f^2 \nabla ^\mu (f^{-1}\nabla_\mu B)- \Omega_\mu \nabla^\mu
 E +\alpha f\nabla_\mu \phi \nabla^\mu B &=0 \,,
\label{Bianchi_decomp}
\end{align}
where we have used an abbreviation
\begin{align}
\Omega_\mu  =2(E\nabla_\mu B-B \nabla_\mu E)\,,
\label{Omega}
\end{align}
which corresponds to the twist of $V^\mu $, i.e., 
$\Omega_\mu =\epsilon _{\mu\nu\rho\sigma }V^\nu \nabla^\rho V^\sigma $ .
Equation~(\ref{Omega}) manifests that the supersymmetric solution 
can be rotating only in the dyonic case.

At this stage we introduce a local coordinate system. 
Since $V$ is Killing $\nabla_{(\mu }V_{\nu)}=0$,
the most desirable choice is  
$V^\mu =(\partial/\partial t)^\mu $ for 
which the metric components are independent of the time coordinate $t$.  
Thus, the spacetime metric can be locally written as  a 
twisted fibre bundle over the 3-space as 
\begin{align}
\D s^2=- f (\D t+\omega  ) ^2 +f^{-1} h_{mn}\D x^m \D x^n   \,, 
\label{time_metric2}
\end{align}
where $f^{-1}h_{mn}$ $(m,n,...=1,2,3)$ 
is the metric of the orbit space of the action of $V$. 
The 1-form $\omega $ corresponds to the rotation of $V$, which measures the 
gravito-electromagnetic Sagnac connection. 
Viewing $V =-f(\D t+\omega )$, equation (\ref{dV_timelike}) gives
the governing equation for $\omega $ as 
\begin{align}
\nabla_{[\mu }\omega_{\nu ] }=\frac{1}{2f^2}\epsilon_{\mu\nu\rho\sigma
 }V^\rho \Omega^\sigma \,,\label{domega}
\end{align}
which determines $\omega $ uniquely 
modulo a gradient of a scalar function.

Besides, there exists a local scalar $z$ such
that $a_\mu=\nabla_\mu z$ due to $\D a=0$. 
Let $D_m $ denote the covariant derivative associated with 
the base space metric $h_{mn}$. Equation~(\ref{derv_a}) then 
implies that $z$ is harmonic  $D^2 z=0$, so that  we can use $z$
as one of the coordinate on the base space. 
Thus, the 3-metric $h_{mn}$ may be decomposed as 
\begin{align}
h_{mn}\D x^m \D x^n =\tilde h_{MN}(\D x^M+k^M \D
 z)(\D x^N+k^N\D z ) +\D z^2 \,, 
\label{time_metric1}
\end{align}
where the indices $M, N,...$ range from 1 to 2 with 
$x^1=x$ and $x^2=y$.

Observe that the above metric form has a large degrees of gauge freedom. 
One may easily deduce that the metric is invariant under the change of coordinate
\begin{align}
 t \to t-\lambda (x^m) \,, \qquad {\rm and}  \qquad \omega  \to \omega
 +\D \lambda (x^m)
 \,,
\label{coord1}
\end{align}
which is the gauge transformation of the Kaluza-Klein gauge field
$\omega $. This freedom will be used to eliminate the integration function 
arising from~(\ref{domega}), so we  
remain it unspecified at present. 
In addition the coordinate transformation $x^M= x^M(x'^N, z)$ 
is permissible. 
Using this freedom we can always choose the coordinates $x^M$ in such a way that
\begin{align}
x^M =x^M (x'^N, z)\,, \qquad {\rm with} \qquad 
\frac{\partial x^M}{\partial z}=-k^M\,, 
\end{align}
which eliminates the vector $k^M$ from the metric. 
In the following discussion we can, without loss of no generality, 
restrict the 3-metric $h_{mn}$ to take the form, 
\begin{align}
h_{mn}\D x^m \D x^n =\tilde h_{MN}(x, y, z)\D x^M \D x^N+ \D z^2 \,.
\label{3metric}
\end{align}
We shall refer to this 3-dimensional Riemannian manifold 
as a ``base space.''

Let us turn to examine~(\ref{da_timelike}). 
Define a projection operator 
\begin{align}
{\tilde h^\mu }_{~\nu} =f {\delta ^\mu }_\nu +V^\mu V_\nu -a^\mu
 a_\nu \,, 
\end{align}
which can be regarded as    
$\tilde h_{\mu\nu }=\tilde h_{MN}(\nabla_\mu x^M)(\nabla_\nu x^N)$.
The nonvanishing components of~(\ref{da_timelike}) boil down to 
\begin{align}
{\tilde h ^\rho}_{~\mu}{\tilde h^\sigma}_{~\nu} \nabla _\rho a_\sigma 
=-\tfrac 12 {\tilde h}_{\mu\nu }a^\rho \nabla_\rho f \,.\label{totally_geo}
\end{align}
We can view this equation as such that the level set $z={\rm constant}$ is a totally
geodesic submanifold with respect to the base space 
$\tilde h_{MN}\D x^M\D x^N +\D z^2$, i.e., its extrinsic curvature is zero.
This requires that $\tilde h_{MN}$ is independent of the coordinate $z$,  
$\partial_z \tilde h_{MN}=0$.

We next investigate the dilatino equation~(\ref{dil_con}), which 
are divided into two cases depending on $\alpha \ne 0$ or $\alpha=0$. 
In the following subsections we shall examine these cases separately.

\subsubsection{The $\alpha\ne 0$ case}

Inspecting~(\ref{derv_E}), (\ref{dil_con_31}),
(\ref{derv_B}) and (\ref{dil_con_41}), 
one finds
\begin{align}
D _m   (Ee^{-\phi/\alpha })=0 \,, 
\qquad 
D_m (Be^{\phi/\alpha })=0 \,.
\label{E_sol1}
\end{align}
These are easily solved as 
\begin{align}
 E=c_{\rm E} e^{\phi/\alpha }\,, \qquad B=c_{\rm B} e^{-\phi/\alpha }\,,
\label{EB_sol}
\end{align}
where $(c_{\rm E}, c_{\rm B})$ are constants. 
Taking note of a useful relation
\begin{align}
 D_m\phi -\frac{\alpha }{2f}D_m \left(E^2-B^2\right)=0 \,,
\label{dphi}
\end{align}
one can find that all other dilatino 
equations~(\ref{dil_con}) are satisfied. 
Unlike the ordinary supergravities, 
we must check the dilaton field equation so as to 
keep the consistency with the dilatino equation. 
Substitution of (\ref{EB_sol}) into~(\ref{dilaton}) yields 
\begin{align}
D_m \left(f^{-(1+\alpha^2)/2}D^m  \phi \right)=0 \,,
\label{dilaton_sol}
\end{align}
Indices $m, n ,...$ are raised and
lowered by $h_{mn}$ and its inverse.   
The above equation (\ref{dilaton_sol}) is to be compared with the 
equations for the gauge fields below.

Substituting~(\ref{EB_sol}),  equations (\ref{Maxeq_decomp}) and 
(\ref{Bianchi_decomp}) simplify to 
\begin{align}
c_{\rm E} \left[ D^2 \phi +\left\{
\frac{1-\alpha^2 }{\alpha }-\frac{2(E^2-3 B^2)}
{\alpha (E^2+B^2)}
\right\} (D \phi )^2 \right]=&0
 \,, \label{phisol1}\\
c_{\rm B} \left[ D^2 \phi -\left\{
\frac{1-\alpha^2 }{\alpha }+\frac{2(3 E^2-B^2)}{\alpha (E^2+B^2)}
\right\} (D \phi )^2 \right]
=&0 \label{phisol2}\,. 
\end{align}
Consider first the case  $c_{\rm E}c_{\rm B}\ne 0$ where the 
solution is dyonic ($F_{\mu\nu }\star F^{\mu\nu }\ne 0$). The above two equations   
yield 
\begin{align}
(3-\alpha^2 )(D \phi )^2 =0 \,, \label{phisol3}
\end{align}
and 
\begin{align}
 D^2\left[ \left(c_{\rm E}^2 e^{4\phi/\alpha }+c_{\rm B}^2 \right)^{-1}\right]=0 \,.\label{phisol4}&
\end{align} 
For the generic coupling ($\alpha\ne \sqrt 3$), 
equation~(\ref{phisol3}) implies that the supersymmetric dyonic solution has a  
constant dilaton field, whence $E=B={\rm constant}$. 
Since the Maxwell field vanishes in the constant dilaton case
[see~(\ref{F_timelike})],  
this is nothing but a vacuum supersymmetric solution, 
i.e.,  the Minkowski spacetime.

In the dyonic case, 
equation~(\ref{phisol3}) shows that 
a nontrivial dilaton arises only for $\alpha =\sqrt 3$, which corresponds
to  the Kaluza-Klein compactification of 5-dimensional vacuum
gravity.~\footnote{
The Kaluza-Klein coupling exhibits well behavior since the positive energy
theorem be shown in arbitrary dimensions. This is shown in appendix.} 
Indeed equations~(\ref{dilaton_sol}) and ~(\ref{phisol4}) are compatible if and
only if $\alpha=\sqrt 3$, 
as expected from~(\ref{integrability_dil}). 
Furthermore, in the dyonic case, only the $\alpha=\sqrt 3$ case  is 
consistent with the integrability condition of (\ref{domega}):
\begin{align}
\nabla_\mu \left(f^{-2} \Omega^\mu \right)=0\,. 
\label{intebrability_domega}
\end{align}
Tod used this condition as a consistency condition and 
obtained the general solutions~\cite{Tod2}.

From~(\ref{F_timelike})  the dilaton is given by  
\begin{align}
  \phi = \frac{\sqrt 3}{4}\ln \left(\frac{c_{\rm E}^2+c_{\rm B}^2-c_{\rm B}^2 H}{c_{\rm E}^2
 H}\right) \,, \label{dyon1}
\end{align}
where $H$ stands for a harmonic function on the base space $D^2 H=0$. 
The norm $f$ of a Killing field and 
the rotation form $\omega $~(\ref{domega}) are successively obtained as 
\begin{align} 
f &= \frac{c_{\rm E} (c_{\rm E}^2+c_{\rm B}^2)}{\sqrt{H (c_{\rm E}^2+c_{\rm B}^2-c_{\rm B}^2 H)}} \,, \qquad 
\partial_{[m }\omega_{n]}=\frac{c_{\rm B}}{2c_{\rm E}(c_{\rm E}^2+c_{\rm B}^2) }{}^{(h)} \epsilon_{mnp}
D^p H \,,
\label{N_omega}
\end{align}
where  
${}^{(h)} \epsilon_{mnp} $ is the volume-element compatible with the
3-metric $h_{mn} $ of the base space (\ref{3metric}) with 
$-V\we {}^{(h)} \epsilon $ being positively oriented. 
From (\ref{phisol4}),  we can find the gauge potential  $F=\D A $, 
\begin{align}
 \qquad 
 A =\frac{c_{\rm E}^2+c_{\rm B}^2}{2c_{\rm E}H}\left(\D t + \omega 
\right) \,.  \label{dyon12}
\end{align}
One can also obtain the corresponding dualized ones~(\ref{duality}):
\begin{align}
\tilde \phi & = -\frac{\sqrt 3}{4}\ln \left(\frac{c_{\rm E}^2+c_{\rm B}^2-c_{\rm B}^2 H}{c_{\rm E}^2 H}\right) \,,
 \qquad 
\tilde A  = -\frac{c_{\rm B}(c_{\rm E}^2+c_{\rm B}^2)(1-H)}
{2 (c_{\rm E}^2+c_{\rm B}^2-c_{\rm B}^2 H)}
\left[\D t+\frac{c_{\rm E}^2 \omega }{c_{\rm B}^2(1-H)}\right]\,,
\label{dyon2}
\end{align}
where $\tilde F=\D \tilde A$.

Although we have introduced two integration constants $(c_{\rm E}, c_{\rm B})$, 
only one of them is of physical relevance.  
To see this consider a scaling the Killing spinor
\begin{align}
 \epsilon ~ \to ~ C \epsilon \,,
\label{scaling}
\end{align}
where $C$ is a complex constant. 
Then the metric~(\ref{time_metric2}) and the Maxwell field~(\ref{F_timelike})
transform as
$f\to |C|^4 f$, $\omega \to |C|^{-4}\omega$ and 
$F \to |C|^{-2}F$, that is to say, we can choose $c_{\rm E}$ or 
$c_{\rm B}$ to take any value we wish. 
The choice $c_{\rm E}={\rm sech}\sigma $ and 
$c_{\rm B}=\tanh\sigma $ ($\sigma \in {\mathbb R}$) is 
physically definitive provided $H$ goes to unity at infinity, 
since $\phi \to 0$ and $f \to 1$ for the above value.

In the purely electric case, i.e., 
$c_{\rm E}\ne 0$ and $c_{\rm B}=0$, one can set $c_{\rm E}=1$ by the 
scaling freedom as described above. In this case, the Bianchi identity
automatically holds and $\alpha $ can take any
value since $F_{\mu\nu }\star F^{\mu\nu } =0$ is satisfied. 
Then we find from~(\ref{phisol1}) that the dilaton and 
the electromagnetic fields are given by
\begin{align}
\phi = -\frac{\alpha }{1+\alpha^2 } \ln H  \,,  
\qquad A = \frac{\D t}{\sqrt{1+\alpha^2 }H}\,.
\label{harmonics_electric} 
\end{align}
Here $H $ is a harmonic function on the base space, 
$D^2 H=0 $. 
For the purely magnetic case, setting $c_{\rm E}=0$ and
$c_{\rm B} \equiv 1$ amounts to the duality
rotation~(\ref{duality}) of the purely electric case:
\begin{align}
\tilde \phi = \frac{\alpha }{1+\alpha^2 } \ln H \,,  
\qquad \partial_{[m}\tilde A_{n]} 
= -\frac 1{2\sqrt{1+\alpha^2 }} {}^{(h)}\epsilon_{mnp}D^p H \,. 
\label{harmonics_magnetic} 
\end{align}
Since either electric field or magnetic field vanishes, 
$\Omega_\mu=0 $ holds, to wit $\D \omega=0$. Hence $\omega $ is locally 
gradient of some scalar function, which can be made to vanish by 
incorporating into the definition of $t$ by exploiting the gauge freedom~(\ref{coord1}).
It follows that $V$ is hypersurface orthogonal and the spacetime is
static for the purely electric/magnetic case.  

Remark that the 2-metric $\tilde h_{MN}(x, y)$ 
is still unrestricted at the current moment.

\subsubsection{The $\alpha =0$ case}

Next, we discuss the $\alpha=0 $ case. 
Contraction of $V^\mu $ to~(\ref{dil_con_51})
gives $\phi={\rm constant}$.  It follows that the
Brans-Dicke-Maxwell system reduces to a usual 
Einstein-Maxwell theory due to supersymmetry. 
Thus its timelike family of supersymmetric
solution is given by the Israel-Wilson-Perj\'es (IWP) 
solution~\cite{Perjes:1971gv}. 
For completeness we shall also discuss this case 
within the present framework, which should recover the result in~\cite{Tod}.  
Let $\Psi =E-{\rm i}B$ denote a complex Ernst-Maxwell
potential~\cite{Ernst:1967wx}.  
Then the Maxwell equations
$\D \star F=0$~(\ref{Maxeq_decomp}) and 
the Bianchi identity $\D F=0$~(\ref{Bianchi_decomp})  
are combined to give the 3-dimensional (complex) Laplace equation 
$ D^2 \Psi ^{-1}=0$. 
Looking at (\ref{Omega}), the solution can be rotating 
only in the dyonic case. The undetermined 2-metric $\tilde h_{MN}$ 
will be found by the integrability condition of the Killing spinor
equation, as demonstrated below.

\subsubsection{Integrability condition}

So far we have been discussed constraints on the geometry and matter
fields which are necessary for the existence of Killing spinor. 
We have exhausted the equations satisfied by bosonic quantities, leaving
the 2-metric $\tilde h_{MN}$ undetermined. 
We shall next look at the Killing spinor equations and   
examine whether further restriction is imposed. 
Adopting the tetrad frame
\begin{align}
e^0 =f^{1/2}(\D t+\omega )  \,, \qquad
 e^I=f^{-1/2}\hat e^I ~~(I=1,2) \,, \qquad e^3 =f^{-1/2}\D z\,,
\end{align}
where $\tilde h_{MN }=\delta_{IJ}\hat e^I_{~M}\hat e^J_{~N}$, 
 equation~(\ref{projection}) reads
\begin{align}
{\rm i}\gamma^0 \epsilon =
f^{-1/2}(E+{\rm i}B\gamma_5) \epsilon \,.
\label{projection2}
\end{align}
Under this condition, 
the time and spatial components of Killing spinor equation are written as 
\begin{align}
\partial_t \epsilon=0 \,, \qquad 
\left[D_m -\omega_m \partial_t +\frac{1}{4f}\left(-\partial_m f+2{\rm
 i}\Omega_m \gamma_5 \right)\right]\epsilon =0 \,,
\label{KSeq_comp}
\end{align}
where we have treated the spatial components at once instead of
discriminating components $x, y $ from $z$.  
The first equation shows that the Killing spinor is time-independent.  
Defining chiral spinors 
\begin{align}
\epsilon^\pm :=\frac{1\pm \gamma_5}{2\sqrt{E \mp{\rm i}B}} \epsilon \,, 
\end{align}
with $\gamma_5\epsilon^\pm =\pm \epsilon^\pm$, 
the second equation of (\ref{KSeq_comp}) reduces to 
\begin{align}
 D_m \epsilon^\pm =0 \,,
\label{cov_con}
\end{align}
viz, $\epsilon^\pm$ are  covariantly constant spinors for the base space. 
It follows that  the 
the solution of the Killing spinor equation is given by 
\begin{align}
\epsilon =\sqrt{E-{\rm i}B} \epsilon ^++ 
\sqrt{E+{\rm i}B} \epsilon^- 
\label{KS_solpm}
\end{align} 
where $\epsilon^\pm$ are the spatially parallel and chiral spinors satisfying 
$\gamma_5 \epsilon^\pm =\pm \epsilon^\pm $.
In the purely electric or magnetic case where $\alpha $ is arbitrary, 
it is further simplified to 
\begin{align}
 \epsilon= H^{-1/[2(1+\alpha^2 )]} \epsilon_{\infty }\,,
\label{KS_sol}
\end{align}  
where $H$ is harmonic (\ref{harmonics_electric}) and 
$\epsilon_{\infty }$ is the spatially parallel spinor
independent of $t$, corresponding to the
asymptotic value of $\epsilon $ and satisfying 
${\rm i}\gamma^0\epsilon_\infty =\epsilon_\infty $. 
It is worth commenting that the condition
${\rm i}\gamma_5\gamma^3\epsilon=f^{-1/2}(E+{\rm i}B\gamma_5)\epsilon$
is not used to derive (\ref{KS_solpm}) and (\ref{KS_sol}).

The integrability condition of (\ref{cov_con}) is
\begin{align}
0=[ D_m ,  D_n]\epsilon^\pm = \frac{1}{2}\left(
h_{m[p}{}^{(h)}S_{q]n}-h_{n[p}{}^{(h)}S_{q]m} \right)\gamma^{pq}\epsilon^\pm \,,
\end{align}
where we have replaced the Riemann tensor by the Schouten
tensor 
$^{(h)} S_{mn}:={}^{(h)} R_{mn}-(1/4){}^{(h)} Rh_{mn}$
for the 3-metric $ h_{mn}$. Contracting with $\bar \epsilon $ 
and $\bar \epsilon \gamma_5 $, 
we obtain 
\begin{align}
 {^{(h)} S^p}_{[m}{\Phi_{n]p}}=0 \,, \qquad 
  {^{(h)} S^p}_{[m}\star {\Phi_{n]p}}=0 \,. 
\label{Int_KS_IJ}
\end{align}
Combined with~(\ref{Phi_timelike}), (\ref{3metric}) and~(\ref{totally_geo}), 
we come to the conclusion that the base space~(\ref{3metric})  is Ricci
flat (${}^{(h)} R_{mn}=0$), thence flat since it is 3-dimensional. 
This means that the spacetime is
conforma-stationary~\cite{Stephani:2003tm} and 
($\epsilon^\pm, \epsilon_\infty $) are constant spinors.
We can also find that the dilatino equation~(\ref{dilatino}) 
is satisfied automatically under the projection~(\ref{projection2}). 
Since equation~(\ref{projection2}) is the only restriction,  
the solution preserves at least half of supersymmetries.

We have only solved the gravitino and dilatino Killing equations,  
the dilaton equation of motion,  the Maxwell equations 
and the Bianchi identity. We have nowhere used 
Einstein's equations, but they automatically 
hold as an integrability condition for the Killing spinor equation.
From~(\ref{KS}),  we get 
\begin{align}
 \nabla_{[\mu }\nabla_{\nu] }\epsilon =\frac{1}{8}R_{\mu\nu\rho\sigma
 }\gamma^{\rho\sigma }\epsilon 
=-\frac{{\rm i}}{4\sqrt{1+\alpha^2 }}
\gamma^{\rho\sigma }\gamma_{[\nu} 
\nabla_{\mu]} (e^{-\alpha\phi }F_{\rho\sigma})\cdot \epsilon
 -\frac{e^{-2\alpha\phi }}{16 (1+\alpha^2 )}\gamma^{\rho\sigma}
\gamma_{[\nu }\gamma^{\lambda\tau }\gamma_{\mu ]}F_{\rho\sigma}
F_{\lambda\tau}\epsilon \,. 
\end{align}
Contracting $\gamma^\nu $ to this equation and 
{\it using the dilatino equation}~(\ref{dilatino}) and the first 
Bianchi identity $R_{\mu [\nu\rho\sigma ]}=0$, 
we find 
\begin{align}
\left[
\ma E_{\mu\nu }\gamma^\nu +\frac{\rm i}{2 \sqrt{1+\alpha^2
 }}e^{-\alpha\phi }\left\{\gamma^{\nu\rho\sigma }
\gamma_\mu \nabla_{[\nu }
 F_{\rho\sigma] }-2e^{2\alpha\phi }\gamma_\nu \gamma_\mu \nabla_\rho 
(e^{-2\alpha\phi }F^{\nu\rho })\right\}
\right]\epsilon =0 \,, 
\label{integrability_KS}
\end{align}
where  we have defined  
\begin{align}
 \ma E_{\mu\nu}:=R_{\mu\nu }-2(\nabla_\mu\phi)
(\nabla_\nu\phi)-T_{\mu\nu}^{(\rm em)}
 \,. 
\end{align}
Here, $\ma E_{\mu\nu}=0$ is equivalent to Einstein's equations~(\ref{Einsteineq}).
From (\ref{integrability_KS}), when the Bianchi identity and the
Maxwell equations for $F$ are satisfied, we deduce that
$\ma E_{\mu\nu }\gamma^{\nu }\epsilon =0 $. 
Contracting with $\bar \epsilon $, one finds
\begin{align}
 \ma E_{\mu\nu }V^\nu =0 \,. 
\label{E0i}
\end{align} 
If we dot it with $\ma E_{\mu \rho }\gamma^\rho $, we get
\begin{align}
 \ma E_{\mu\nu }{\ma E_\mu}^\nu =0 \,, 
\qquad \textrm{(no sum on $\mu $)} \,. 
\label{Eij}
\end{align} 
In the orthonormal frame, equation~(\ref{E0i}) 
implies $\ma E_{00}=\ma E_{0i}=0$ where 
$i,j,..=1,2,3$ and (\ref{Eij}) 
implies $\ma E_{ij}=0$, as we desired to show. 
This has been already demonstrated in (\ref{Int_KS_IJ}).

\subsubsection{Summary}

Let us encapsulate the results in this section. 
The timelike family of 
supersymmetric solutions in Einstein-Maxwell-dilaton 
system where the dilatino equation implies the 
dilaton equation are either of the followings: 
\begin{itemize}
\item[(i)] 
{\it A dyonic and rotating solution for $\alpha=\sqrt 3$}: 
the metric is written as a conforma-stationary form 
\begin{align}
 \D s^2=-f\left(\D t+\vec \omega \cdot \D \vec x\right)^2 +f^{-1}\D \vec x^2
 \,. \label{time_metric3}
\end{align}
where $H$ is harmonic on the base space $\vec \nabla^2H=0$, 
$f$ and $\vec \omega $ are given by~(\ref{N_omega}) and the
dilaton and the gauge fields are~(\ref{dyon1}), (\ref{dyon12}) or (\ref{dyon2}). 
The solution of the Killing spinor equation is given by~(\ref{KS_solpm})
where $E$ and $B$ are given by~(\ref{EB_sol}). 

\item[(ii)]  
{\it A purely electric or magnetic static solution for arbitrary
$\alpha $}: 
the spacetime is the Gibbons-Maeda solution~\cite{GM},
\begin{align}
\D s^2 =- H ^{-2/(1+\alpha^2 )} \D t^2 +H ^{2/(1+\alpha^2 )}\D
\vec x^2 \,, \label{GM}
\end{align} 
where the dilaton and the gauge fields are given
by~(\ref{harmonics_electric}) for the electric case and
	     by~(\ref{harmonics_magnetic}) for the magnetic case. 
The solution of the Killing spinor equation is given by (\ref{KS_sol}).

\item[(iii)] 
{\it A dyonic and rotating solution for $\alpha=0$}:
this reduces to the BPS solution in the Einstein-Maxwell system and
described by the IWP metric~\cite{Perjes:1971gv}, 
\begin{align}
\D s^2 =-|\Psi|^2 (\D t+\vec \omega \cdot \D \vec x )^2 +|\Psi|^{-2}
\D \vec x^2\,,
\label{IWP}
\end{align}
where $\Psi$ is a complex harmonic function $\vec \nabla^2 \Psi =0$ and 
$\vec \omega $ is given by quadrature~(\ref{domega}) as 
$\vec\nabla\times\vec\omega=\I(\bar\Psi^{-1}\vec\nabla\Psi^{-1}-\Psi^{-1}\vec\nabla\bar\Psi^{-1})$.
The solution of the Killing spinor is given by~(\ref{KS_solpm}) with 
$E$ and $B$ obeying $D^2 \Psi^{-1}=0$.  
\end{itemize}

Except for the Majumdar-Papapetrou solution [the static solution in case
(iii)], these solutions do not describe black hole spacetimes.

\subsection{Null family}
\label{sec:null}

In this section we study the case in which 
$V^\mu $ is null, i.e.,  $E=B=0$. 
The Maxwell field $F_{\mu\nu }$ and $\Phi_{\mu\nu }$ satisfy
\begin{align}
&i_V F=0 \,, \qquad i_V\star  F=0 \,, \qquad 
 i_V \Phi =0 \,, \qquad i_V\star  \Phi =0 \,, \nonumber \\   
&\Phi_{\mu\nu }\Phi^{\mu\nu }=0 \,, \qquad \Phi_{\mu\nu }\star \Phi^{\mu\nu
 }=0\,,  \qquad  {\Phi_{(\mu }}^\rho \star \Phi_{\nu )\rho }=0\,.
\label{FPhi_null}
\end{align}
These relations are sufficient to establish 
\begin{align}
F_{\mu \nu }F^{\mu \nu }=0 \,, \qquad 
F^{\mu\mu} \star F^{\mu \nu}=0 \,, \qquad 
F_{\mu\nu }\Phi^{\mu\nu }=0 \,, \qquad
 F_{\mu\nu }\star \Phi^{\mu\nu }=0\,.
\label{FPhi_null2}
\end{align}
As opposed to the timelike case, 
$F_{\mu\nu }\star F^{\mu\nu }=0$ always holds for the null case.  
We are thus not concerned with the dilaton field equation since it 
is ensured by dilatino equation.  
The dilatino equation~(\ref{dil_con}) imposes a single restriction 
\begin{align}
V \we \D \phi =0\,.
\label{Vwephi}
\end{align}

Equation~(\ref{derv_V}) means that 
the vector field $V^\mu $ is covariantly-conserved 
$\nabla_\mu V_\nu =0$, i.e., the spacetime of the null family describes
a {\it pp}-wave~\cite{Stephani:2003tm}. The $pp$-wave spacetime 
always belongs to the Petrov type N.  
Since $V$ is closed $\D V=0$ and tangent to the affine parametrized geodesic
$V^\mu\nabla_\mu V^\nu =0$, $V$ can be written as  
\begin{align}
V_\mu =-\nabla_\mu  u \,, \qquad 
V^\mu =\left(\frac{\partial }{\partial v}\right)^\mu  \,, 
\end{align}
where $u$ is some scalar function and $v$ is an affine parameter of the
geodesics. Then the metric is independent of $v$
and can be cast into the form~\cite{Stephani:2003tm}
\begin{align}
\D s^2 =-2 \D u\left(\D v+\ma H \D u+\beta _i \D x^i \right)
+ \tilde g_{ij}\D x^i \D x^j\,. 
\label{null_metric}
\end{align}
Here $\ma H$, $\beta _i$ and $\tilde g_{ij}$ ($i, j=1,2$) are functions of 
$u$ and $x^i$. 
Using the coordinate transformation of $x^i$, 
the 2-dimensional metric $\tilde g_{ij}$ 
can be written in a conformally flat form,
\begin{align}
\tilde g_{ij}\D x^i \D x^j =\Omega ^2 \left(\D x^2+\D y^2\right)\,, 
\end{align}
where $\Omega=\Omega (u,x,y)$. 
Equation~(\ref{Vwephi}) now implies that the dilaton is a function only of
$u$,  $\phi =\phi(u)$. One may thence regard the scalar field as a
``null dust'' since the stress-tensor takes the form
$T_{\mu\nu }^{(\phi)}=2(\D \phi/\D u)^2V_\mu V_\nu $.

Due to $a^\mu a_\mu =f=0$, 
the pseudo-vector $a^\mu $ is null or identically
zero. Since $V \we a =0$, 
 there exists a function $\kappa =\kappa (u, x^i)$ such that 
\begin{align}
a_\mu =\kappa V_\mu \,.
\label{akappa}
\end{align} 
From $\D a=0$, one finds $\kappa =\kappa (u)$, hence
$\nabla_\mu \kappa =-[\D \kappa (u)/\D u] V_\mu $.  
It follows that equations~(\ref{derv_a}) and (\ref{derv_Phi}) simplify to 
\begin{align}
\frac{\D \kappa }{\D u}V_\mu  V_\nu  & = \frac{2 e^{-\alpha \phi }}{\sqrt{1+\alpha^2 }}
{F_{(\mu }}^\rho \star \Phi_{\nu )\rho }\,, \label{derv_a2}\\
\nabla_\mu \Phi_{\nu\rho  } &= \frac{\kappa e^{-\alpha \phi } }{\sqrt{1+\alpha^2
 }}
\left(-V_\mu \star F_{\nu\rho }+\epsilon_{\nu\rho \sigma\tau }V^\tau 
 {F_\mu }^\sigma \right)\,.\label{derv_Phi2}
\end{align}

Let us introduce a tetrad frame
\begin{align}
e^+ =\D u\,, \qquad e^- =\D v+\ma H \D u+\beta_i \D x^i \,, \qquad 
e^i =\Omega \D x^i\,,
\label{null_basis}
\end{align}
which obey the orthogonality relations
$\eta _{ab } {e^a}_\mu {e^b}_\nu =g_{\mu \nu }$
with $\epsilon_{+-12}=1$, where 
$\eta _{+-}=\eta_{-+}=-1$, $\eta_{ij}=\delta_{ij}$ and other components vanish. 
Then the condition $i_VF=i_V\star F=0 $ determines the form of Maxwell fields
as
\begin{align}
 F=F_{+i}e^+\we e^i \,, \qquad 
\star F= -\epsilon_{ij}{F_{+i}}e^+\we e^j \,,
\end{align}
where $\epsilon_{12}=-\epsilon_{21}=1$ and the summation over $i,j,...$
is understood.  
Noting $\phi=\phi(u)$, 
the Bianchi identity $\D F=0$ and the Maxwell equation
$\D (e^{-2\alpha\phi }\star F)=0$ reduce to 
\begin{align}
\partial_{[i}\left(\Omega F_{j]+}\right) =0\,, \qquad 
\partial_i \left(\Omega F_{i+}\right)=0 \,.
\end{align} 
It follows that there exists a function 
$\ma F=\ma F (u, x^i)$ such that 
\begin{align}
F_{+i}=-\Omega ^{-1}\partial_i \ma F \,, \qquad \Delta  \ma F =0\,,
\label{fpi_sol}
\end{align}
where $\Delta  \equiv \partial^2/\partial x^2+\partial^2/\partial y^2$.
Thus, $\ma F $ is a harmonic function  
on a flat 2-space $\D x^2+\D y^2$ with a $u$-dependence.

Equation~(\ref{derv_Phi2}) now implies 
\begin{align}
\D \Phi =0 \,, \qquad \D \star \Phi =0\,.
\end{align}
Noticing $i_V\Phi=i_V\star \Phi =0$, we can conclude 
by the parallel argument as above that there exists a harmonic function
$\ma P=\ma P (u, x^i) $ such that 
\begin{align}
\Phi =-\Omega^{-1}\partial_i \ma P e^+\we e^i =-\D u \we \D \ma P \,, \qquad 
\Delta  \ma P =0 \,.
\label{Phi_pot}
\end{align}
Substituting (\ref{Phi_pot}) back into~(\ref{derv_Phi2}), 
we obtain 
\begin{align}
\Omega \partial _i \partial_j \ma P -2\partial_{(i}\ma P 
 \partial_{j)} \Omega &= -\delta_{ij}\partial_k \ma P 
 \partial_k    \Omega \,, \label{derlam}\\
\Omega \partial_u \left(\Omega^{-1}\partial_i \ma P \right)
+\frac{1}{2\Omega^2}W_{ij}\partial_j \ma P &=
\frac{1}{\sqrt{1+\alpha^2 }}e^{-\alpha\phi }\kappa
 \epsilon_{ij}\partial_j \ma F \,,
\label{derlam2}
\end{align}
where 
$W_{ij}:=\partial_i\beta_j-\partial_j \beta_i=(\partial_x\beta_y-\partial_y\beta_x)\epsilon_{ij}$.
Inserting (\ref{fpi_sol}) and (\ref{Phi_pot}) into~(\ref{derv_a2}) gives
\begin{align}
\sqrt{1+\alpha^2 } \frac{\D \kappa }{\D u}e^{\alpha\phi }=2\Omega^{-2}\epsilon_{ij}
\partial_i \ma F \partial_j \ma P \,,
\label{da_eq}
\end{align}
where the left hand side is dependent only on $u$. 
Multiplying $\partial _i\ma P$ to~(\ref{derlam2}) and 
using~(\ref{da_eq}),  we find 
\begin{align}
\partial_u \left[\Omega^{-2}\partial_i\ma P\partial_i\ma P +\frac
 12 \kappa^2 \right]=0\,.
\label{da_eq2}
\end{align}
Equation~(\ref{PhiPhi}) now reduces to
\begin{align}
\Omega^{-2} \partial_i\ma P \partial_i \ma P = 1-\kappa ^2  \,.
\label{PhiPhi2}
\end{align}
Comparing~(\ref{da_eq2}) and (\ref{PhiPhi2}), we arrive at 
$\kappa ={\rm constant}$.

Thus far, we have proceeded in a quite general metric form~(\ref{null_metric}). 
The metric~(\ref{null_metric}) is
invariant under the three kind of coordinate transformations~\cite{Stephani:2003tm}.  
Letting $\zeta =x+{\rm i}y$ and 
$W=(\beta_x-{\rm i}\beta_y)/\sqrt 2$, the metric-form remains intact
under 
$\zeta \to \zeta '=h (\zeta, u)$ with 
\begin{align}
\Omega'^2 =\frac{\Omega^2}{\partial_\zeta h\partial_{\bar \zeta}{\bar
 h}}\,, \qquad W'=\frac{W}{\partial_\zeta h}+
\frac{\Omega^2 \partial_u \bar h }{\partial_\zeta h\partial_{\bar
 \zeta}{\bar h}} \,,\qquad \ma H'=\ma H -\frac{\Omega^2 \partial_u h\partial_u \bar
 h+W\partial_u h\partial_{\bar \zeta }\bar h+\bar W \partial_u
 \bar h\partial_\zeta h}
{\partial_\zeta h\partial_{\bar \zeta}{\bar h}}\,,
\end{align}
where $h$ is analytic in $\zeta $. 
Using the above freedom, we can always  
adopt $\ma P$ as a one of the coordinate of the wave surface as
$\ma P =x$.  
Then, equations~(\ref{derlam}) and (\ref{da_eq2}) imply that 
$\Omega $ is constant, which can be taken as $\Omega \equiv 1$
without losing any generality by means of a simple scaling
$u\to u'= \Omega u $, $v\to v'=\Omega^{-1}v$ and 
$\zeta  \to \zeta ' =\Omega^{-1 }\zeta $ with  
$\ma H' = \Omega^{-2 }\ma H$, 
which also leaves $\Phi$ invariant.  With these choices
($\ma P=x$ and $\Omega =1$),  
$\kappa=0 $ is obtained from~(\ref{PhiPhi2}), i.e., 
$a_\mu =0$.

Equation~(\ref{da_eq}) then leads to $\ma F =\ma F (x ,u)$. 
Since $\ma F $ is harmonic, it is restricted to the form
\begin{align}
\ma F= \ma F_0 (u)x+\ma F _1(u) \,, 
\end{align}
where $(\ma F_0, \ma F_1)$ are arbitrary functions of $u$. 
The function $\ma F_1$ can be gauged away since the Maxwell field
strength $F$ is not affected by
this term. 
From (\ref{derlam2}) we can obtain $W_{ij}=0$, implying that 
$\beta_i$ is a local gradient. This function can be set to zero 
by the transformation $v \to v'=v+g(\zeta, \bar \zeta,  u)$ with
\begin{align}
W'=W-\partial_\zeta g\,,
\qquad \ma H'=\ma H-\partial_u g \,,
\end{align}
which corresponds to the choice of  the $v =0$ surface. 

Finally, the remaining function $\ma H$ can be obtained 
by use of the ($+,+$)-component of Einstein's equation.
Other components of Einstein's equations are 
ensured to hold automatically as an integrability of the Killing spinor
equation. Working in the basis (\ref{null_basis}),  
(\ref{E0i}) implies $\ma E_{-i}=0$ and (\ref{Eij}) implies 
$\ma E_{+i}=\ma E_{ij}=0$, as desired. 
The $(+, +)$-component of the Ricci tensor for the
metric~(\ref{null_metric}) reads
\begin{align}
R_{++}=\frac{1}{2\Omega^4} \left[2\Omega^2 \left(\Delta  \ma
 H-\partial_u \partial_i\beta_i \right)+\frac 12
 W_{ij}W_{ij}-4\Omega^3\partial_u^2 \Omega \right] \,.
\end{align}
Setting $\beta_i=0$ and $\Omega=1$, we arrive at the governing equation for
$\ma H$: 
\begin{align}
\Delta  \ma H (u, x, y) = 2 e^{-2\alpha\phi (u)}\ma F_0 (u)^2 
+2 \left(\frac{\D \phi}{\D u}\right)^2  \,.\label{Poisson}
\end{align}

To sum up, the necessary condition for the supersymmetry in the 
null class requires that the spacetime is  
$pp$-wave~\cite{Stephani:2003tm} described by the metric 
\begin{align}
\D s^2 =-2 \D u \left[\D v +\ma H (u, x, y)\D u \right]
+\D x^2+\D y^2 \,, \qquad 
F= -\ma F _0 (u) \D u \we \D x \,, 
\label{pp_wave}
\end{align}
where $\ma F_0 (u)$ is an arbitrary function characterizing the strength of
the radiative Maxwell field.  $\ma H$ is determined
by the Poisson equation~(\ref{Poisson}) for a given dilaton profile 
$\phi(u)$. Remark that~(\ref{Poisson}) determines $\ma H$ up to another
arbitrary harmonic function $\ma H_0$ with an arbitrary $u$-dependence.

Equation~(\ref{projection}) implies 
\begin{align}
\gamma^+ \epsilon =0\,.
\label{gamp}
\end{align}
Writing out the  
the Killing spinor equation for the metric~(\ref{pp_wave})
and using (\ref{gamp}), we have 
\begin{align}
 \left(\partial_u+\frac{{\rm i}}{\sqrt{1+\alpha^2 }}e^{-\alpha \phi
 (u)}\ma F_0 (u) \gamma^1 \right) \epsilon =0 \,, \qquad 
 \partial_v \epsilon =0 \,, \qquad \partial_i\epsilon =0 \,,
\end{align}
which can be solved as 
\begin{align}
\epsilon =\exp \left[ -\frac{\rm i}{\sqrt{1+\alpha^2 }}\int^u
 \D u e^{-\alpha\phi (u)}\ma F_0 (u)\gamma^1 \right]\epsilon_0 \,, 
\end{align}
where $\epsilon_0 $ is a constant spinor obeying $\gamma^+\epsilon_0=0$.
The dilatino equation imposes no further condition.  
Since the projection (\ref{gamp}) is a unique restriction,  
the solution preserves at least half of supersymmetries.

\section{Novel properties of BPS solutions}
\label{sec:pp}

We explore some characteristic properties of supersymmetric solutions
obtained in the previous section. 
This issue has not been addressed in~\cite{Tod}.
The following subsection 
enumerates all the maximally supersymmetric solutions. In the next two 
subsections, we study several aspects of BPS solutions from the
viewpoints of conserved charges, sigma models and the 
Kaluza-Klein embedding.   
The dyonic solution in the 
timelike family is not entirely new, since it can be generated by 
the 5-dimensional transformations.

\subsection{Maximal supersymmetry}

The maximally supersymmetric solutions in this
theory can be obtained as follows. 
To restore the complete supersymmetries, the dilatino
equation must impose no algebraic  constraints. This means that terms in the
basis 
$\{{\bf 1}, \gamma_5, \gamma_\mu ,\gamma_\mu \gamma_5, \gamma_{\mu\nu }\}$  
of the gamma matrix must vanish separately. We are then led to 
\begin{align}
 \phi=\phi_0\,, \qquad F_{\mu\nu }=0 \,,
\end{align} 
for $\alpha\ne 0$ and 
$\phi=\phi_0$ for $\alpha=0$. 
The $\alpha\ne 0$ case is then tantamount to the vacuum case, so that the maximally
supersymmetric solution is only the Minkowski spacetime. 
For $\alpha=0$, the maximally supersymmetric solutions in
Einstein-Maxwell theory are obtained, which are the Minkowski spacetime, 
the Nariai-Bertotti-Robinson spacetime ${\rm AdS}_2\times S^2 $~\cite{Nariai}, 
\begin{align}
 \D s^2 = -\frac{r^2}{Q^2} \D t^2+ \frac{Q^2}{r^2}\D r^2+Q^2 \left(\D
 \theta ^2+\sin^2\theta \D \phi ^2 \right)\,, \qquad 
F=Q^{-1} \D t\we \D r \,,
\end{align}
where $Q$ is a constant corresponding to the Maxwell charge, 
and 
the Kowalski-Glikman $pp$-wave~\cite{KowalskiGlikman:1985im},
\begin{align}
 \D s_2^2 =-2\D u \left[\D u+ \tfrac 12\lambda^2 (x^2+y^2)\D u\right]+\D x^2+\D
 y^2 \,, \qquad 
F= \lambda \D u \we \D x \,,
\end{align} 
where $\lambda $ is a constant. 
All of these backgrounds are  conformally flat $C_{\mu\nu\rho\sigma }=0$ and 
the Maxwell field is covariantly constant $\nabla_\mu F_{\nu\rho }=0$.

\subsection{Force balance}

Each BPS solution is specified by a single harmonic function $H$ on a
flat base space,  which is taken to be the multi-center point sources 
$H= 1+\sum_k q^{(k)}/|\vec x-\vec x_{(k)}|$.
The Gibbons-Maeda metric~(\ref{GM}) with $\alpha\ne 0$ 
is asymptotically flat, and saturates the BPS inequality~(\ref{BPS}). 
It describes the collection of naked singularities, 
instead of the multiple configuration of black holes. A single charge
cannot anchor the black hole to have a nonvanishing horizon.  
The IWP family ($\alpha=0$) also describes naked singularities with the exception
of the Reissner-Nordstr\"om solution~\cite{Hartle:1972ya}.

The dyonic solution~(\ref{time_metric3}) is not singular at the point sources 
$\vec x=\vec x_{(k)}$, but they 
do not correspond to the locus of horizons since the circumferential radius vanishes there.  
Furthermore the dyonic solution is not asymptotically flat in the strict sense due to the
NUT charge, thereby   
the solution fails to satisfy the BPS bound~(\ref{BPS}). 
Instead, the spacetime is asymptotically locally flat, wherein 
a NUT charge plays an interesting r\^ole as a gravitational dyon.
Letting $(r, \theta ,\varphi)$ be the spherical coordinates at infinity, 
we shall define (in an appropriate gauge) the scalar charge $\Sigma $ and
the NUT charge $N$ by
\begin{align}
 \phi \sim \pm \frac{\Sigma }{r}\,, \qquad 
g_{t\varphi}  \sim 
\pm 2Ng_{tt}\cos\theta  \,,
\end{align} 
as $r\to \infty $. 
It can be easily verified  that the dyonic solution~(\ref{time_metric3}) 
with $c_{\rm E}={\rm sech}\sigma $ and $c_{\rm B}=\tanh\sigma $
satisfies  the ``anti-gravity condition'' of
Scherk~\cite{Scherk:1979aj},  
\begin{align}
 M^2 +\Sigma^2 +N^2 =Q_e^2+Q_m^2 \,,
\label{antigravity}
\end{align}
where $M$, $Q_e$ and $Q_m$ have been read off from the ``monopole terms'' for
the metric and the gauge potential. 
This equation just encodes the superposition principle, which
is distinguished from the BPS condition~(\ref{BPS})  
expressed only in terms of global charges.

Let us consider an additional implication of the relations between 
(\ref{BPS}) and (\ref{antigravity}).  
From~(\ref{em_H}) and~(\ref{dV_timelike}), 
the electromagnetic parts of the Nester 2-form can be rewritten 
as $H_{\mu\nu }=2\nabla_\mu V_\nu $, thence its integral gives 
\begin{align}
M_{\rm BPS}=-\frac{1}{2} \int_{\partial \Sigma }\D S_{\mu\nu
 }H^{\mu\nu } =-\int_{\partial \Sigma }\nabla^\mu V^\nu \D S_{\mu\nu
 }=M_{\rm Komar}\,. 
\label{Komar}
\end{align}
This accords precisely with the expression of Komar integral for the timelike
Killing vector $V^\mu $~\cite{Komar:1958wp}. 
It follows that the failure of the saturation of the BPS inequality~(\ref{BPS}) 
stems from the disagreement of the ADM mass and the Komar charge. 
This is of course outside the reach of supersymmetry, which is
essentially local whilst the conserved charge in the gravitating
system is a global notion. 
If the spacetime is asymptotically flat in the usual sense, 
the ADM mass and the Komar energy coincide $M=M_{\rm Komar}$, as expected.

Since the timelike family of solutions is necessarily 
stationary, it is also enlightening to discuss the  
relation to the non-BPS, stationary Einstein-Maxwell-dilaton system,  
which dimensionally reduces to the gravity-coupled sigma
model. The sigma model analysis will reveal that BPS solutions occupy a
distinguished position compared to non-BPS solutions.

A spacetime in Einstein-Maxwell-dilaton gravity 
admitting a timelike group of motions generated by a 
Killing vector $V^\mu $ with norm $V^\mu V_\mu=-f(<0) $ is described by the action,
\begin{align}
S_3 & = \int \D ^3 x \sqrt{h}
\left[{}^{(h)}R - \mas G_{AB}(\Phi^C) h^{mn}
(D_m\Phi^A)(D_n\Phi^B)\right]\,. 
\label{NLSM}
\end{align}
where $f^{-1}h_{mn}$ is the metric orthogonal to the orbits of $V$
as~(\ref{time_metric2}) and ${}^{(h)} R$ is the Ricci curvature
of $h_{mn}$. 
Here, $\Phi^A$ ($A=1, ...,5$) constitutes the five real scalars~\cite{Galtsov:1995mb}
\begin{align}
 \Phi^A =(f , \psi,  v, a ,\phi) \,,
\end{align} 
where  $\phi$ is a dilaton and
\begin{align}
\partial_m v = {\sqrt 2} F_{m\mu }V^\mu \,, \qquad 
\partial_m a = - \sqrt 2 e^{-2\alpha\phi }\star F_{m \mu }V^\mu \,, \qquad 
\partial_m \psi = \Omega_m -(v\partial_ma -a\partial_m v)\,,    
\end{align}
with $\Omega =-\star (V\we \D V)$ being the twist of $V$. 
The target space metric $\mas G_{AB}$ reads~\cite{Galtsov:1995mb}
\begin{align}
\mas G_{AB}\D \Phi^A \D \Phi^B = 
\frac{1}{2 f^2 }\left[\D f^2 +(\D \psi +v\D a -a\D v)^2\right]
-\frac{e^{-2\alpha\phi }\D v^2+e^{2\alpha\phi }\D a^2}{f}+2\D \phi ^2
 \,,
\label{target}
\end{align}
which is symmetric  iff $\alpha=0, \sqrt 3$ and  
Einstein iff $\alpha=\sqrt 3$. The Euler-Lagrange equations 
derived from the action~(\ref{NLSM}) define a harmonic map from the 
base space to the target space.

Comparing with the timelike class of supersymmetric solution, 
the above scalars take the form, 
\begin{align}
f=E^2+B^2  \,, \qquad 
\psi =0 \,, 
\end{align}
and 
\begin{align}
E= c_{\rm E} e^{\phi/\sqrt 3}\,, \qquad 
B= c_{\rm B}e^{-\phi/\sqrt 3} \,, \qquad 
v= \frac{c_{\rm E}}{\sqrt 2}e^{4\phi/\sqrt 3}\,, \qquad 
a =\frac{c_{\rm B}}{\sqrt 2}e^{-4\phi/\sqrt 3}
\,,
\end{align}
for the dyonic case, and 
\begin{align}
E= e^{\phi/\alpha }\,, \qquad 
B= 0 \,, \qquad 
v= \sqrt{\frac{2}{1+\alpha^2 }} e^{[(1+\alpha^2 )/\alpha]\phi} \,,
 \qquad 
a= 0 \,,
\end{align}
for the purely electric case.  The purely magnetic case is obtained by
$E\leftrightarrow B$,
$a\leftrightarrow v$ and $\phi\leftrightarrow -\phi$. 
In every case, the supersymmetric solutions 
correspond to the null geodesics of the target space
$\mas G_{AB}\D \Phi^A \D \Phi^B=0$ with a harmonic being its affine 
parameter.  
Since the target space metric acts as a source of 3-dimensional Euclidean
gravity~(\ref{NLSM}), this implies that the 3-metric $h_{mn}$ is flat,
which appears to be responsible for producing a state of 
equipoise~\cite{Rasheed:1995zv,Clement:1986bt,Chemissany:2010zp}.

Incidentally, the multiple solution~(\ref{Gibbons}) is not described by 
null geodesics of the target space aside from the Majumdar-Papapetrou
solution ($H_1=H_2$, i.e, $\phi=0$ and $F_{\mu\nu }F^{\mu \nu }=0$).  
We can deduce that this may also due to~(\ref{integrability_dil}).   
It is therefore reasonable to infer that there exist other 
 multi-soliton solutions which are not described
by null geodesics on the target space~(\ref{target}). 
Moreover, there indeed exist multi-center solutions that are described
by null geodesics on the target space~(\ref{target}) but not the BPS solutions to the 
Einstein-Maxwell-dilaton gravity. In order to gain further insight into 
equilibrium solutions, the analysis of all geodesics is required.       
Unfortunately the approach given in~\cite{Rasheed:1995zv,Clement:1986bt,Chemissany:2010zp} 
seems inapplicable since the sigma-model representation on coset spaces
has been fully exploited wherein.  
The direct evaluation of geodesics for the space~(\ref{target}) 
seems more promising for this purpose. The interrelation among 
between BPS solutions and equilibrium states is somewhat obscure and 
deserves further detailed investigation. We hope to visit this issue
elsewhere.

\subsection{Liftup to 5-dimensional BPS solutions}

Since the dyonic solution has $\alpha=\sqrt{3}$, 
the solution can be uplifted into five-dimensional vacuum gravity 
via the Kaluza-Klein ansatz, 
\begin{align}
\D s_5^2 &=e^{-4\phi/\sqrt 3} \left(\D x^5+2 A_\mu \D x^\mu  \right)^2
 +e^{2\phi/\sqrt 3} g_{\mu\nu }\D x^\mu \D x^\nu \,.
\label{KK}
\end{align}
We discuss the supersymmetric solutions with $\alpha=\sqrt 3$ 
obtained in the previous section from the 5-dimensional perspective. 
The BPS solutions in Einstein-Maxwell-dilaton gravity with 
$\alpha=\sqrt 3$ should constitute a subset of 5-dimensional 
vacuum BPS solutions with a spatial isometry.

The timelike family of BPS solutions for 5-dimensional vacuum gravity is 
static\footnote{
Setting $F=0$ in (3.3) of reference~\cite{GGHPR} leads to 
$f={\rm constant}$, $G^+=G^-=0$. Hence 
 $\omega=0 $ is concluded. 
} and given by the direct product of a flat time-direction and 
a hyper-K\"ahler manifold~\cite{GGHPR}, 
\begin{align}
 \D s^2 =- \D t^2+\D s_{\rm HK}^2 \,.
\end{align} 
As a hyper-K\"ahler manifold, 
we choose the Gibbons-Hawking space~\cite{Gibbons:1979zt}
\begin{align}
 \D s^2=h^{-1} \left(\D x^5 +\vec \chi \cdot \D \vec x\right)^2 
+h\D \vec x^2 \,, \qquad \vec \nabla \times \vec \chi =\vec \nabla h \,.
\label{GH}
\end{align}
Here the integrability condition of the equation for $\chi$ implies that 
$h$ is harmonic on $\mathbb R^3$. The vector field
$\partial/\partial x^5$ is a triholomorphic Killing vector which 
preserves the three complex structures invariant. The Gibbons-Hawking
space naturally  leads to dimensional reduction~\cite{GGHPR}. 
Then, the 5-dimensional metric reads
\begin{align}
 \D s^2_5 & =-\D t^2 + h ^{-1} \left(\D x^5 +\vec \chi \cdot \D \vec x\right) ^2 
+h \D \vec x ^2 \,, \qquad 
\vec \nabla \times \vec \chi =\vec \nabla h \,,
\label{GPS}
\end{align} 
where the metric is independent of 
$t$ and $x^5$. $h$ is harmonic on the flat 3-dimensional space 
$\D \vec x^2$ and $\partial/\partial x^5$ preserves the 3 complex structures.   
This metric  
describes a (multiple generalization of) Gross-Perry-Sorkin monopole~\cite{GPS}. 
Compactifying along $x^5$ via the ansatz~(\ref{KK}), 
we find that the 4-dimensional Einstein metric $g_{\mu\nu }$ 
is the magnetic Gibbons-Maeda solution with $\alpha=\sqrt{3}$.

Applying the Lorentz boost along ($t, x^5$)-plane,
\begin{align}
 \D t ~ \to ~ \D t \cosh \sigma  +\D x^5 \sinh \sigma   \,,\qquad 
 \D x^5 ~\to ~ \D x^5 \cosh\sigma  +\D t\sinh \sigma  \,,
\end{align}
where $\tanh \sigma  $ controls the boost velocity,  
we obtain a rotating metric from~(\ref{GPS}),  
\begin{align}
\D s_5^2 =& \left(h^{-1}\cosh^2\sigma  -\sinh^2\sigma  \right)\left[\D x^5 +
\frac{\cosh\sigma \sinh\sigma  (1-h) }{\cosh^2\sigma  -h\sinh^2\sigma }
\left(\D t+\frac{\vec \chi \cdot \D \vec x}{(1-h)\sinh\sigma  }\right)\right]^2
\nonumber \\
& -\frac{1}{\cosh^2\sigma  -h\sinh^2\sigma }\left(\D t+\sinh\sigma  \vec
 \chi \cdot \D \vec x
 \right)^2+h \D \vec x^2 \,.
\end{align}
The dimensional reduction gives the dyonic supersymmetric
solution~(\ref{N_omega}) and (\ref{dyon2})  with $H=h$, 
$c_{\rm E}={\rm sech} \sigma  $ and $c_{\rm B}=\tanh\sigma  $.

The Kaluza-Klein embedding can be applied for the null case as well.
The general null class of 5-dimensional vacuum BPS solution is 
the $pp$-wave~\cite{GGHPR}, 
\begin{align}
\D s^2_5 =-2 \D u \left[\D v+\ma H(u,\vec x) \D u \right] +
\left[\D \vec x +\vec x\times \vec \omega(u) \D u\right]^2 \,, \qquad 
\vec \nabla ^2 \ma H=0 \,,
\label{5d_ppwave}
\end{align}
where we have included for convenience the cross-term $\D u \D \vec x $,
which can be made to vanish by the isometry of 3-dimensional Euclid
space $\D \vec x^2$.  Turning off the $u$-dependence, 
compactification along $u$ with $\vec \omega=0$ gives rise to the 
electrically charged Gibbons-Maeda solution~(\ref{GM})~\cite{Gibbons:1982ih}. 
Applying the Lorentz boost in the ($v, z$)-plane simply generates 
gauge transformation and does not alter the 4-dimensional solution. 

In order to obtain the 4-dimensional $pp$-wave geometry~(\ref{pp_wave}) form~(\ref{5d_ppwave}), 
consider a coordinate transformation 
\begin{align}
&\D u =\Omega_1(u')^2 \D u'  \,, \qquad x=\Omega _1(u')x'\,, \qquad 
 y=\Omega_1 (u') y' \,, \qquad 
 z =\Omega_3(u') z' \,, \nonumber \\
& v =v'+\frac 12\left[\Omega_1^{-1}\dot \Omega_1 (x'^2+y'^2)
+\Omega_1^{-2}\Omega_3\dot \Omega_3 z'^2\right]-\Omega_1\Omega_3\omega_2'x'z'\,,
\end{align}
with $\omega' _1=\omega'_3=0$ and 
$\omega'_2 =\Omega_1(u')^{-5 }\ma F_0(u')$, where 
$\vec \omega'(u')=\vec \omega(u)$. The dot denotes the derivative 
with respect to $u'$.
Then the 5-dimensional metric~(\ref{5d_ppwave}) translates into
\begin{align}
 \D s_5^2 = \Omega_1(u')^2 \left[-2 \D u'\left(\D v'+\ma H'\D u' \right)
+\D x'^2+\D y'^2  \right] 
+\Omega_3(u')^2 \left[\D z'+2 \ma F_0(u') x' \D u'\right]^2 
\end{align}
where 
\begin{align}
 \ma H' = \Omega_1^{2}\ma H-\frac{1}{2\Omega_1^3}
\biggl[&(x'^2+y'^2)(-\Omega_1\ddot\Omega_1+2\Omega_1\dot \Omega_1^2 )
+\omega_2^2\Omega_1^5 (-3x'^2\Omega_1^2+z'^2\Omega_3^2)
\nonumber \\
& 
+z'^2\Omega_3 (2\dot \Omega_1\dot \Omega_3-\Omega_1\ddot \Omega_3)
+2x'z'\Omega_1^4 (\Omega_3\dot \omega_2 +2\omega_2 \dot \Omega_3 )
\biggl] \,.
\end{align}
Since it is always possible to choose $\ma H'$ to be independent of 
coordinate $z'$, the dimensional reduction along $z'$ gives the desired 
metric (\ref{pp_wave}) by taking
$\Omega_3=\Omega_1^{-2}=e^{-2\phi(u')/\sqrt 3}$.

\section{Concluding remarks}
\label{sec:summary}

We investigated the supersymmetric solutions in 4-dimensional
Einstein-Maxwell-dilaton theory with an arbitrary coupling constant
$\alpha $. The primary motivation to examine this theory comes from the  
fact that properties of static (nonextremal) black hole solutions 
are very sensitive to the coupling constant, and that the 
rotating black hole solution has not been found yet.
In the light of sigma model, 
the target space metric becomes homogeneous only for 
$\alpha =0, \sqrt 3$, in which a coset representative is possible. 
For other values of $\alpha $  the nontrivial transformation is 
unavailable. Still, 
in the case of $\alpha \le \sqrt 3$ the sectional curvature of the
potential space~(\ref{target}) is negative semi-definite, which can be used to prove
the uniqueness theorem of (yet to be discovered) rotating and nonextremal black
holes~\cite{Yazadjiev:2010bj}. 
This encourages us to inquire the extremal limits of these solutions. 
In this paper, we studied the supersymmetric solutions satisfying the 
gravitino and the dilatino Killing spinor equations.

The supersymmetric Killing spinor equations were derived~in~\cite{GKLTT} 
together with the Bogomol'nyi bound. However, it has been 
known that  the equilibrium solution~(\ref{Gibbons}) fails to satisfy the 
Bogomol'nyi bound although the coupling constant is the value 
inspired by string theory.    
In this paper we reverted back to the 1st-order Killing spinor equations
and found that the dilatino equation does not imply the dilation field
equation except for $\alpha=0, \sqrt 3$, which correspond respectively
to the Brans-Dicke-Maxwell theory  and the
Kaluza-Klein reduction of 5-dimensional vacuum gravity, and 
$F_{\mu\nu }\star F^{\mu\nu }=0$ for which the solution is purely
electric or magnetic.  
Otherwise,  the Einstein-Maxwell-dilaton gravity would not be embedded
into supergravity theory.  We may attribute this to the fact 
that the axion field resulting from the 10 dimensional heterotic string
theory cannot be truncated consistently unless $F_{\mu\nu }\star F^{\mu\nu }=0$. Hence 
the static dyonic multiple solution~(\ref{Gibbons}) is not the BPS
solution to the Einstein-Maxwell-dilaton gravity, although it enjoys
the superposition principle.  This is also related to the fact that the 
solution~(\ref{Gibbons}) does not have the null geodesic description 
on the target space. 
The same is true for the dyonic Reissner-Nordstr\"om solution, 
which is given by $H_1=H_2$ in equation~(\ref{Gibbons}) with a
trivial dilaton. 
Since this metric fulfills $F_{\mu\nu }F^{\mu\nu }=0$, 
it is an exact solution in the Einstein-Maxwell-dilaton system.     
Though, it is not the supersymmetric solution to this theory since
it does not satisfy the dilatino equation 
despite being mechanical equilibrium. 
We should regard it as a BPS solution of the Einstein-Maxwell gravity,
rather than the Einstein-Maxwell-dilaton theory. 

The integrability of the dilatino Killing spinor equation 
also uncovers why only the supersymmetric solutions with 
$\alpha=\sqrt 3$ can be rotating. Although all the 
supersymmetric solutions were obtained in~\cite{Tod2},     
we performed the systematic classification using the modern technique 
and made it clear why this is the case.   
We addressed physical properties of these supersymmetric solutions, 
which have not been addressed in~\cite{Tod2}.  
Looking from 5-dimensions, the dyonic solutions are generated via 
boosting the purely magnetic Gross-Perry-Sorkin monopole solution. 
It has been argued that the nonexistence of multi-spinning configurations may be related to the 
discrepancy of gyromagnetic ratio between the probe particle and the
background spacetime~\cite{Shiromizu:1999xm}. The results in the present
paper are not inconsistent 
with the claim of~\cite{Shiromizu:1999xm} 
since the dyonic metric~(\ref{dyon1}) is not asymptotically flat 
due to the NUT charge.  
Unfortunately, all the solutions in the timelike family do not 
describe regular black holes. It should be noted that this does not mean the nonexistence of  
nonextremal rotating dilatonic black holes.   
For the null family, 
the dilatino equation automatically ensures the dilaton equation of
motion provided the Maxwell equations and the Bianchi identity are satisfied. 
Both families of solutions preserve at least half of supersymmetries. 
The full restoration of supersymmetries occurs only for the Minkowski
spacetime for $\alpha\ne 0$.

The present work can be extended into several directions.
We expect that the appropriate incorporation of an axion field 
will give rise to the correct square-root equation for an arbitrary coupling.    
The result~\cite{Nozawa:2010zg} strongly implies that the  
the theory~(\ref{action}) can be ``gauged'' to include the exponential
Liouville-type potential. It is interesting to see whether the gauged
dilaton gravity admits a Bogomol'nyi-type inequality. 
The classification of pseudo supersymmetric solutions in dilatonic ``fake supergravity'' 
also seems to be a plausible generalization to the present work.

\acknowledgements

The author would like to thank 
Dietmar Klemm, Jun-ichirou Koga and Kei-ichi Maeda  for 
stimulating discussions. 
This work was partially supported by the MEXT Grant-in-Aid for
Scientific Research on Innovative Areas No. 21111006

\appendix

\section{Bogomol'nyi bound in arbitrary dimensions}
\label{app:BPS}

In this appendix we shall consider 
the $d$-dimensional Einstein-Maxwell-dilaton action coupled to matter
fields, 
\begin{align}
 S= \frac{1}{16\pi G_d} \int \D ^d x \sqrt{-g}
\left[R-2(\nabla ^\mu \phi)(\nabla_\mu \phi )-e^{-2\alpha\phi} F_{\mu\nu
 }F^{\mu\nu }\right] +S_{\rm matter} \,,
\label{action_ddim}
\end{align}
where $S_{\rm matter}$ describes the action for the matter fields.
The Einstein-Maxwell-dilaton equations are 
\begin{align}
G_{\mu\nu } &=T_{\mu\nu }^{(\rm em)}+T^{(\phi)}_{\mu\nu }+
8\pi G_d T_{\mu\nu
 }{(\rm mat.)} \,, \\
\nabla_\nu (e^{-2\alpha \phi }F^{\mu\nu })&=4\pi G_d J^\mu ({\rm mat.}) \,, \\
\nabla_\mu \nabla^\mu \phi +\frac \alpha 2 e^{-2\alpha \phi}F_{\mu\nu
 }F^{\mu\nu }&=-4\pi G_d \rho ({\rm mat.})\,, 
\end{align}
where 
\begin{align}
T_{\mu\nu }^{(\phi)} = 2 \left[ (\nabla_\mu \phi)( \nabla_\nu \phi)
 -\frac 12 g_{\mu\nu }(\nabla_\rho \phi )(\nabla^\rho \phi ) 
\right] \,, \qquad
T_{\mu\nu }^{(\rm em)} = 2 e^{-2\alpha\phi }
\left(F_{\mu \rho }{F_\nu }^\rho -\frac 14 g_{\mu\nu }F_{\rho\sigma
 }F^{\rho\sigma }\right) \,,
\end{align}
and 
\begin{align}
T_{\mu\nu }({\rm mat.}) =\frac{-2}{\sqrt{-g}}\frac{\delta S_{\rm
 matter}}{\delta g^{\mu\nu }} \,, \qquad 
J^\mu ({\rm mat.}) = \frac{1}{\sqrt{-g}}\frac{\delta S_{\rm
 matter}}{\delta A_{\mu }} \,, \qquad 
\rho ({\rm mat. })= \frac{1}{\sqrt{-g}}\frac{\delta S_{\rm matter} }{\delta
 \phi }\,.
\end{align}

When the coupling constant $\alpha $ takes a special value,  
\begin{align}
 \alpha=\alpha_{\rm c}:=\sqrt{\frac{2(d-1)}{d-2}} \,.
\label{KK_ddim}
\end{align} 
the theory~(\ref{action_ddim}) arises via the Kaluza-Klein compactification
of $D=(d+1)$-dimensional vacuum gravity. More precisely,  
a $D$-dimensional spacetime admitting the one-dimensional isometry group 
generated by the spacelike Killing vector 
$\xi =\partial/\partial x^{D}$ can be written
as
\begin{align}
\D s_D^2 =e^{2a\phi} \left(\D x^D+2A_\mu \D x^\mu \right)^2 
+e^{2b\phi}g_{\mu\nu }\D x^\mu \D x^\nu \,,
\label{KK_reduction}
\end{align} 
with
\begin{align}
 a=-\sqrt{\frac{2(d-2)}{d-1}} \,, \qquad 
b =\sqrt{\frac{2}{(d-1)(d-2)}} \,,
\end{align}
where
$g_{\mu\nu }$ is the $d$-dimensional Einstein-frame metric, 
$A_\mu $ and $\phi$ correspond to the twist and the norm of $\xi $, respectively. 
Then the $D$-dimensional Ricci scalar density accords with the 
Lagrangian density in the action~(\ref{action_ddim}) 
up to the total divergence.  So we can identify $A_\mu $ as a 
${\rm U}(1) $ gauge field and $\phi$ as a dilaton field in $d$-dimensions.

We demonstrate the 
Bogomol'nyi inequality in $d$-dimensional Einstein-Maxwell-dilaton theory 
by focusing on the case of Kaluza-Klein coupling, 
$\alpha=\alpha_{\rm c}$. The $d=4$ case seems special, in which 
the theory admits a Bogomol'nyi inequality for an arbitrary coupling
$\alpha $.  
We denote the $d$-dimensional gamma matrix by $\Gamma_\mu $. 
In this section we shall work with a Dirac spinor $\epsilon $.

We define a super-covariant derivative $\hat \nabla _\mu $ 
acting on a complex spinor as 
\begin{align}
\hat \nabla_\mu \epsilon = \left[\nabla_\mu +\frac{\I}{4 (d-2)}e^{-\alpha_{\rm c}\phi}F^{\nu\rho }
\left(\Gamma_\mu \Gamma_{\nu\rho }-2(d-2)g_{\mu[\nu }\Gamma_{\rho]
 }\right)\right]\epsilon \,.
\label{KS_d}
\end{align}
and a variation of dilatino by
\begin{align}
\delta \lambda =\frac{1}{\sqrt 2}\left(\Gamma^\mu \nabla_\mu \phi
-\frac{\I}{4}\alpha_{\rm c}
e^{-\alpha_{\rm c} \phi
 }\Gamma^{\mu\nu }F_{\mu\nu } \right)\epsilon \,.
\label{KS2_d}
\end{align} 
The specific factors appearing in these definitions 
come from the $(d+1)$-dimensional gravitino variations for vacuum
gravity, as well as to give a positivity bound.

In terms of a super-covariant derivative (\ref{KS_d}), 
we define a Nester-Witten tensor
\begin{align}
 \hat E^{\mu\nu }=-{\rm i}
\left(\bar \epsilon \Gamma^{\mu\nu\rho }\hat \nabla_\rho \epsilon 
-\overline{\hat \nabla _\rho }\Gamma^{\mu\nu\rho }\epsilon \right)\,. 
\label{Nester}
\end{align}
Observe that 
$\hat E^{\mu\nu }$ decompose into 
\begin{align}
\hat E^{\mu\nu }=E^{\mu\nu }+H^{\mu\nu }\,,
\end{align}
where $E^{\mu\nu }$ is an ordinary Nester-Witten tensor and 
$H^{\mu\nu }$ denotes the electromagnetic contribution, 
\begin{align}
H^{\mu\nu }=-e^{-\alpha_{\rm c}\phi }\left(
 \bar \epsilon \epsilon F^{\mu\nu} 
+\frac 12 \bar \epsilon \Gamma^{\mu\nu\rho\sigma }
\epsilon F_{\rho\sigma }\right) \,. 
\label{em_contribution}
\end{align}

Consider an asymptotically flat spacetime to which 
an ADM $d$-momentum can be assigned~\cite{Witten:1981mf,ADM}.
This means that  the $d$-dimensional spacetime has to 
admit the spin structure.  
Choose a spatial hypersurface $\Sigma $ with a 
future-pointing unit normal $n^\mu $ and let 
 $\partial \Sigma $ be its boundary at spatial infinity.
Assume that $\epsilon $ asymptotes to a constant spinor
$\epsilon_\infty $,  the dilaton falls off
to zero, and $F_{\mu\nu }$ is purely electric at spatial infinity.  
As in 4-dimensions, 
we find  
\begin{align}
-\int _\Sigma \D \Sigma n_\mu \nabla_\nu \hat E^{\mu \nu } &= 
\frac 12 \int_{\partial \Sigma }\D S_{\mu\nu }\hat E^{\mu\nu }
 \nonumber \\
&= \frac 12 
\int _{\partial\Sigma }\D S_{\mu\nu } E^{\mu\nu }
 -\frac{1}{2}\int _{\partial\Sigma } \D
 S_{\mu\nu }  \bar \epsilon_\infty \epsilon_\infty F^{\mu\nu }
\nonumber \\ 
&=-{\rm i}\bar \epsilon_\infty \gamma^\mu \epsilon_\infty P_\mu
 -\frac{1}{2}\bar \epsilon _\infty \epsilon_\infty Q  \,,\label{Gausseq_d}
\end{align}
where $\D S_{\mu\nu }$ is the element of ($d-2$)-sphere at infinity.
$P_\mu $ denotes the ADM
$d$-momentum~\cite{Witten:1981mf,ADM} and 
\begin{align}
Q = \int_{\partial\Sigma } \D S_{\mu\nu }F^{\mu\nu } \,,
\end{align}
is the total electric charge.

We move on to evaluate the volume integral of (\ref{Gausseq_d}). 
A lengthy calculation reveals that 
\begin{align}
 \nabla_\nu \hat E^{\mu\nu}=&2{\rm i}
\overline{\hat \nabla_\rho \epsilon }\Gamma ^{\mu \nu \rho }
\hat \nabla_\nu \epsilon +2\I  \overline{\delta \lambda }
\Gamma^{\mu }\delta \lambda 
+8\pi G_d K^\mu \,,
\end{align}
where we have imposed the Bianchi identity $\D F=0$ and 
used the abbreviation 
\begin{align}
K^\mu := -{T^\mu }_\nu ({\rm mat.}) (\I \bar \epsilon \Gamma ^\mu
 \epsilon) -\tfrac 12 \bar \epsilon \epsilon e^{\alpha_{\rm c}\phi}
J^\mu({\rm mat.}) \,.
\label{Kmu}
\end{align}
It follows that 
\begin{align}
-{\rm i}\bar \epsilon_\infty \gamma^\mu \epsilon_\infty P_\mu
 -\frac{1}{2}\bar \epsilon _\infty \epsilon_\infty Q =
-\int _\Sigma \D \Sigma n_\mu \left(
2{\rm i}
\overline{\hat \nabla_\rho \epsilon }\Gamma ^{\mu \nu \rho }
\hat \nabla_\nu \epsilon +2\I  \overline{\delta \lambda }
\Gamma^{\mu }\delta \lambda 
+8\pi G_d K^\mu 
\right) \,. \label{Gausseq2}
\end{align}
The first two terms in the integrand on the right-hand side of~(\ref{Gausseq2})
are non-negative for $\epsilon $ satisfying
the modified Dirac-Witten equation 
$ \gamma^{a} \hat D_{a} \epsilon =0 $.
The last term on the right-hand side of (\ref{Gausseq2})
is non-negative if $K^\mu $ is future-directed causal vector. 
Henceforth  we shall assume that this is the case.  
Under these conditions 
the left-hand side of~(\ref{Gausseq2}) has to have non-negative
eigenvalues, giving rise to a desired inequality
\begin{align}
M\ge \tfrac{1}{2} Q \,,  
\label{BPS_d}
\end{align}
where $M=\sqrt{-P_\mu P^\mu}$ is the $d$-dimensional ADM mass. 
The inequality is saturated if and only if the 
asymptotically flat spacetime admits a spinor satisfying
\begin{align}
 \hat \nabla_\mu \epsilon =0 \,, \qquad \delta \lambda =0 \,, \qquad
 K^\mu =0\,.
\end{align}
These are the supersymmetry transformations and derivable from the 
$(d+1)$-dimensional vacuum supergravity.


\end{document}